\documentclass[%
 reprint,
superscriptaddress,
amsmath,amssymb,
aps,
prd,
floatfix,
]{revtex4-2}

\usepackage{graphicx}
\usepackage{dcolumn}
\usepackage{bm}
\usepackage{float}
\usepackage{color}
\usepackage{soul}

\DeclareUnicodeCharacter{2212}{-}
\DeclareUnicodeCharacter{02BC}{'}

\begin{document}

\preprint{APS/123-QED}

\title{Exploring Strangeness Enhancement and Particle Production in Small Collision Systems with EPOS4 at \(\sqrt{s_{\mathrm{NN}}} = 5.02\)~TeV}

\author{Hirak Kumar Koley}
\email{hirak.koley@gmail.com}
\affiliation{Nuclear and Particle Physics Research Centre, Department of Physics, Jadavpur University, Kolkata - 700032, India}

\author{Subikash Choudhury}
\email{subikash.choudhury@gmail.com}
\affiliation{Nuclear and Particle Physics Research Centre, Department of Physics, Jadavpur University, Kolkata - 700032, India}

\author{Argha Deb}
\email{argha$_$deb@yahoo.com}
\affiliation{Nuclear and Particle Physics Research Centre, Department of Physics, Jadavpur University, Kolkata - 700032, India}
\affiliation{School of Studies in Environmental Radiation and Archaeological Sciences, Jadavpur University, Kolkata 700032, India}

\author{Mitali Mondal}
\email{mitalimon@gmail.com}
\affiliation{Nuclear and Particle Physics Research Centre, Department of Physics, Jadavpur University, Kolkata - 700032, India}
\affiliation{School of Studies in Environmental Radiation and Archaeological Sciences, Jadavpur University, Kolkata 700032, India}

\date{\today}

\begin{abstract}

The observation of collectivity and strangeness enhancement in small collision systems, such as proton-proton (pp) and proton-lead (p-Pb) collisions, challenges traditional assumptions regarding thermalization and particle production mechanisms. In this study, we investigate particle yields and transverse momentum distributions in pp and p-Pb collisions at $\sqrt{s_\text{NN}} = 5.02 $ TeV using the EPOS4 event generator, which employs a core-corona framework to model particle production across a variety of system sizes. EPOS4 successfully reproduces many qualitative trends observed in experimental data, including the hardening of $p_{\rm{T}}$-spectra with multiplicity, the hierarchical strangeness enhancement in strange-to-pion ratios, and characteristic modifications of particle yield ratios as a function of $p_{\rm{T}}$ and multiplicity. The microcanonical approach to core hadronization used in EPOS4 seems to provide a more realistic description of small systems compared to grand-canonical treatments. Nonetheless, quantitative discrepancies still persist in describing several physical observables. Future model refinements, including improved core-corona balancing, differential freeze-out conditions for multi-strange hadrons, and incorporation of finite strangeness correlation volumes, may be taken into account for enhancing EPOS4’s predictive power and deepening our understanding of the complex dynamics governing the particle production in high-energy collisions.

\end{abstract}

\maketitle


\section{\label{sec:level1}Introduction}

Recent results from proton-proton (pp) and proton-nucleus (p-A) collisions at LHC energies have revealed several intriguing features that are characteristic of quark-gluon plasma (QGP) \cite{Shuryak_2009} - a dense medium of deconfined quarks and gluons (partons) typically produced in the relativistic collisions of heavy nuclei. 
Notably, the flow-like phenomena observed in these so-called small collision systems suggest the onset of collective dynamics \cite{Nagle_2018, grosseoetringhaus2024}, which in the conventional understanding of such systems, was not anticipated \cite{pikaprlamks0_alice, Khachatryan_2017}. Besides collective flow, small collision systems also show signs of strangeness enhancement \cite{pp_Nature}.

Strangeness enhancement was one of the earliest proposed signatures of QGP formation, as strange quarks, and consequently, strange hadrons, can be thermally produced from the deconfined partonic medium at lower energy thresholds \cite{Koch1986ud, PhysRevLett.48.1066}. Data from CERN-SPS and RHIC support this idea, with strangeness enhancement appearing evident when comparing the relative abundances of strange particles (primarily with pions) in central heavy-ion collisions to those observed in p-Be and/or pp collisions \cite{ANDERSEN1998209, Abelev_2008}. 

Theoretically, the hadronic abundances measured in heavy-ion collisions, including those of (multi-)strange particles, follow the trends predicted by statistical thermal model calculations using the grand canonical ensemble (GCE) approach \cite{Cleymans_2006, ANDRONIC2009516}. 
These models demonstrate better agreement with experimental data as the collisions become more central and the energy increases, suggesting that the produced medium moves toward achieving thermal equilibration. In contrast, strangeness production in peripheral and pp collisions has been found to be significantly suppressed compared to thermal model expectations based on the GCE approach \cite{Hagedorn1984uy}. 
This suppression is understood, at least in part, as a finite volume effect known as canonical suppression, which makes the GCE approach invalid for small systems \cite{tounsi2001, Tounsi_2003}. For the GCE to be applicable, the system volume must be large enough to satisfy the condition VT$^{3}>$1 \cite{RAFELSKI1980279}, where volume V=R$^{3}$ and T is temperature. To address this limitation, a canonical formalism is adopted to extend the scope of thermal model calculations to small systems, where a strict requirement of local strangeness number conservation is imposed for the production of (multi-)strange particles \cite{becattini1997universalitythermalhadronproduction, Redlich_2009}. 
However, this approach necessitates different freeze-out volumes to describe the abundances of strange and non-strange hadrons simultaneously and it struggles to predict the system size dependence of the $\phi$-meson \cite{Acharya_2019, Kraus_2007, Kraus_2009}.

System size remains a crucial factor in determining whether a collision system can achieve thermalization. Femtoscopic measurements have shown that the system sizes of the highest multiplicity p$-$Pb events are actually small, and are comparable to minimum bias pp collisions \cite{Adam_2015}. In fact, the absence of jet quenching in small systems indirectly suggests that the medium produced is not large or dense enough to stop high-momentum particles \cite{Abelev_2013}. Consequently, the question of whether a thermalized medium can be produced, even in the highest multiplicity p$-$Pb events, continues to be a subject of debate.

In QCD-inspired models, such as PYTHIA \cite{Sj_strand_2008}, HERWIG \cite{Bellm_2020}, etc, where thermalization is not assumed, strangeness production at high transverse momentum follows hard partonic interactions like flavour creation, flavour excitations and gluon splitting. 
For low-momentum strangeness production, string fragmentation mechanisms are typically used, where the production of strange hadrons is suppressed relative to non-strange ones due to the heavier constituent mass of the strange quark compared to up or down quarks. However, with these standard implementations of strangeness production, none of these models can capture the complete picture of strangeness enhancement observed in the experimental data, even qualitatively, unless additional phenomenological processes, such as color ropes \cite{bierlich2015}, baryon junctions \cite{Christiansen_2015}, or color reconnection beyond leading colors, are included \cite{Gieseke_2018}. 
Despite incorporating these additional processes, QCD-inspired models still fall short in describing certain aspects of collective flow in high multiplicity pp and p$-$Pb data, such as long-range correlations and the mass ordering of elliptic flow, among others. 

A QGP-inspired model like EPOS has been successful in describing the features of collective flow and strangeness enhancement in data with reasonable accuracy. The model creates a two-component phase based on the local density of strings produced from the initial interactions, consisting of a dense QGP-core and a dilute corona \cite{CoreCorona_Werner, PhysRevC.81.029902}. It mimics the evolution of particle yields and spectra in data as a function of volume or multiplicity by balancing the relative contributions from the core and the corona, respectively. This approach allows EPOS to capture the dynamics of small and large systems more effectively, offering insights into both the collective behavior and the production of strange hadrons \cite{Pierog_2015}.

In this work, we employ EPOS4 \cite{PhysRevCWarner, EPOS4_pp_PbPb, PhysRevC.108.034904, PhysRevC.109.034918}, an upgraded version of EPOS3 \cite{Werner_2014}, to study the transverse momentum spectra, yields, and particle ratios for non-strange, strange, and multi-strange particles in minimum bias as well as as a function of charged particle multiplicity in pp and p-Pb collisions at $\sqrt{s_{\rm NN}} =$ 5.02 TeV. Through a detailed quantitative comparison between data and model calculations, we aim to assess the efficacy of the model, particularly in explaining overall particle production, including strangeness enhancement, in small collision systems. 
A key aspect of EPOS4 is its implementation of a core-corona approach, which enables the model to dynamically account for the transition between independent particle production and collective behavior as a function of local energy density. This feature may offer valuable insights into the limiting charged-particle multiplicity above which QGP-like effects—such as collective flow and strangeness enhancement—begin to manifest in small systems. Given that results from small collision systems are often compared with those from heavy-ion collisions to probe the existence and properties of the QGP, such insights are essential for disentangling initial-state effects from genuine QGP signatures, thereby contributing to a more robust and unambiguous understanding of QGP formation in different collision environments.

The organization of the paper is as follows. In the next section, a detailed description of the EPOS4 model is presented, highlighting the major changes and improvements compared to earlier versions. Section III discusses the results, presenting model predictions for various physical observables alongside comparisons with available experimental data. Finally, the paper concludes with a summary of the key findings and their interpretation in terms of the relevant physics processes implemented in EPOS4.

\section{Event Generator: EPOS4}

EPOS4 is the latest version in the EPOS series, which has been significantly upgraded by implementing parallelization of multiple partonic or nucleonic scatterings, replacing the previous sequential treatment. The EPOS model in general, employs a perturbative QCD (pQCD) inspired Gribov-Regge multiple parton scattering framework \cite{DRESCHER200193}, wherein each individual scattering produces a longitudinally stretched colored flux tube (string) with transverse kinks originating from initial hard scatterings. These flux tubes subsequently fragment into smaller string segments that eventually hadronize into final state particles through a mechanism of independent string fragmentation. When a dense environment of such overlapping string segments are formed- corresponding to a high multiplicity event, it prevents string segments from hadronizing independently. Instead a core-corona approach is adopted whereby a string segment or as it is called pre-hadron becomes the part of core or corona based on an energy loss criterion. String segments exceeding a model-dependent fractional energy loss threshold form the bulk matter or “core” and the rest constitute the “corona”.

Once the core pre-hadrons are identified, corresponding energy-momentum tensor, T$^{\mu\nu}$ and flavor flow vector N$^\mu$$_{q}$ are calculated at a space point $x$  and at initial proper time $\tau_{0}$. Starting from $\tau_{0}$, the core part of the system is subjected to 3+1D viscous hydrodynamic evolution with the choice of shear viscosity to entropy density ratio, $\eta /s = $ 0.08 \cite{Karpenko_2014, Werner_2010}. The hydrodynamic evolution continues until the system reaches a constant energy density hadronization hypersurface defined by $\epsilon_{H}$=0.57 GeV/fm$^{3}$, which corresponds to a temperature of 167 MeV. Hadronization in EPOS4 occurs via a microcanonical statistical hadronization approach, ensuring exact conservation of baryon number, electric charge, and strangeness over the entire hypersurface. It is observed that to maintain the fixed energy density at the hadronization hypersurface, core droplets with higher energy or invariant mass correspond to larger effective volumes, a characteristic of high-multiplicity events. In this limit, microcanonical and grand canonical ensembles produce nearly identical momentum spectra across particle species. In contrast, low-multiplicity events involve smaller core droplets with restricted phase space for microcanonical sampling, leading to a natural suppression of multistrange hadron production, deviating significantly from GCE expectation but in line with experimental observations.

The corona pre-hadrons are the ones which are produced along the periphery of the system or those with sufficient energy to escape the medium with some energy loss. The corona pre-hadrons then hadronize independently via the usual fragmentation processes. All final-state hadrons, from both core and corona, subsequently undergo hadronic re-scattering, modelled using the Ultra-relativistic Quantum Molecular Dynamics (UrQMD) \cite{UrQMD} framework.

\begin{figure}[t]
	\centering 
	\includegraphics[width=0.48\textwidth, height=0.39\textheight, angle=0]{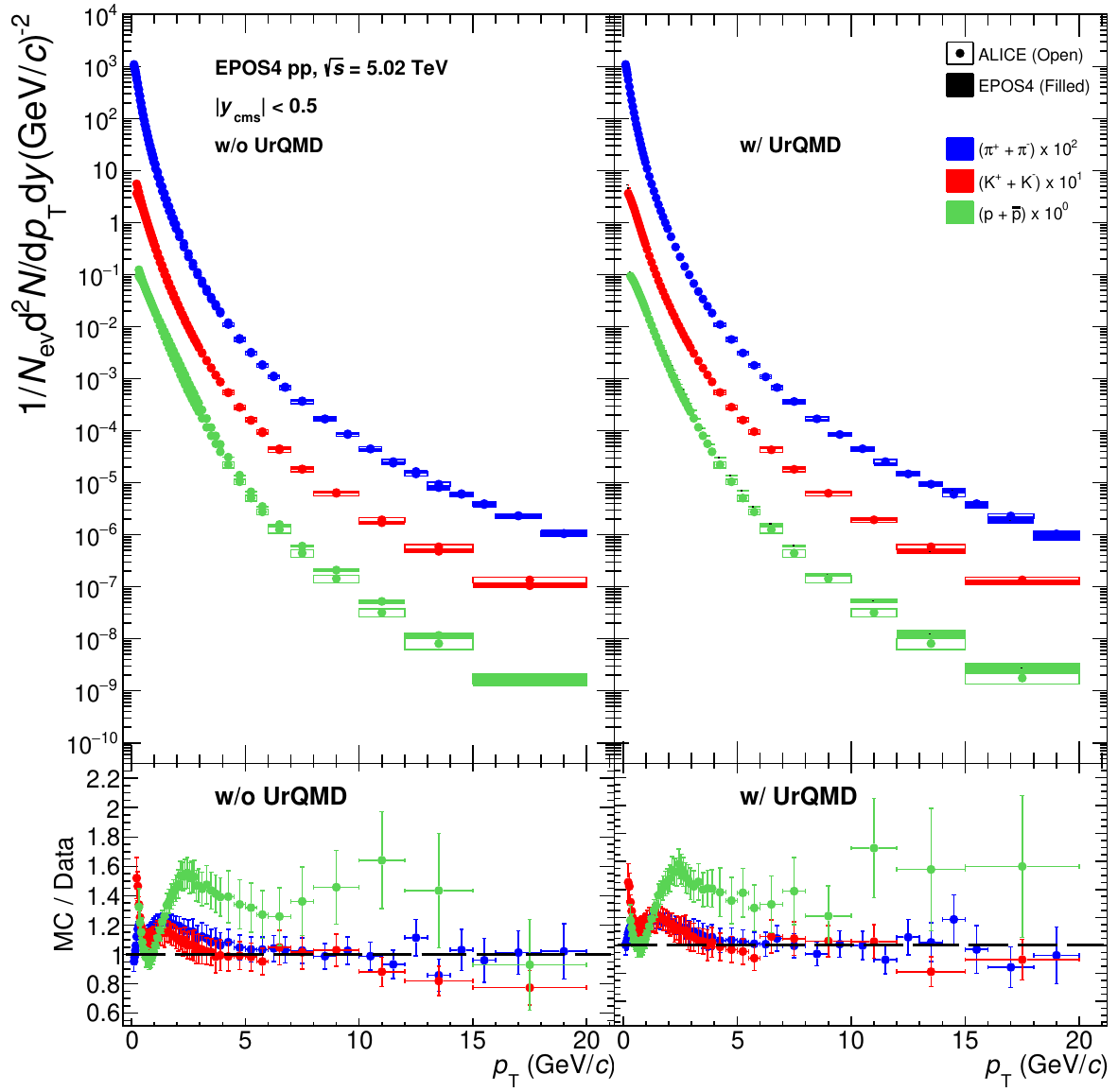}	
 \includegraphics[width=0.48\textwidth, height=0.39\textheight, angle=0]{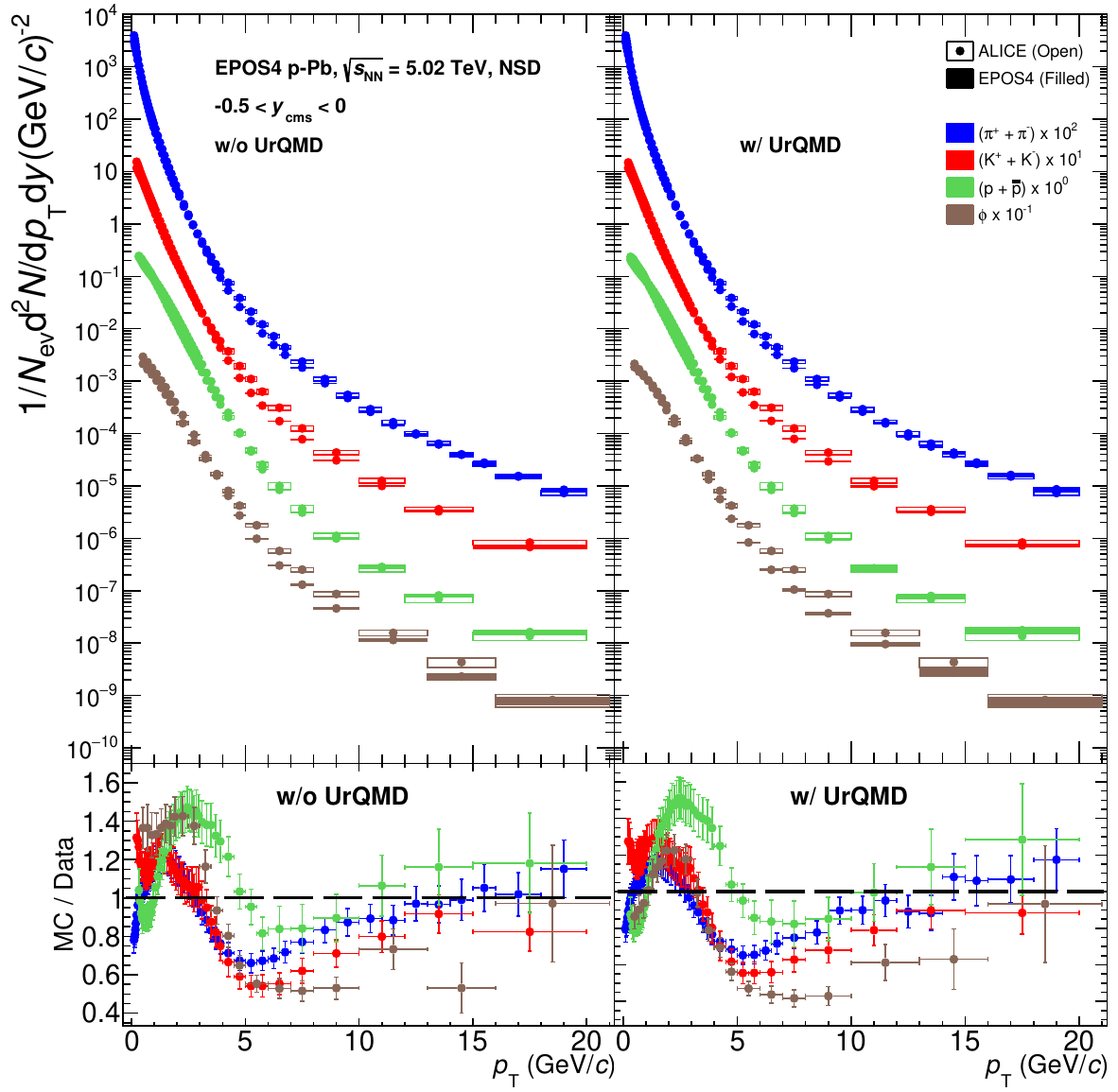}	
	\caption{(Color online) The invariant $p_{\rm{T}}$-differential spectra of $\pi$, K and p in pp collisions at $\sqrt{s}$ = 5.02 TeV (top) and including $\phi$ in p-Pb collisions at $\sqrt{s_{\rm NN}}$ = 5.02 TeV (bottom), compared to ALICE data \cite{ALICE_piKpHighpT, Phi_ALICE, ALICEPikp_pp}} 
	\label{comparisonptspectra}%
\end{figure}

For this study, we have generated approximately 20 million minimum bias events for both pp and p–Pb collisions at $\sqrt{s_{\rm NN}}$ = 5.02 TeV using the EPOS4 event generator. Simulations were performed both with and without the inclusion of the UrQMD hadronic afterburner to account for final-state hadronic interactions. Throughout the text and in figures, samples incorporating the UrQMD afterburner are denoted as “w/ UrQMD”, while those without are referred to as “w/o UrQMD”. Since several physics observables have been studied separately for the core and corona components, it is important to note that core–corona separation is only feasible for EPOS-generated samples without the UrQMD afterburner. By design, EPOS assigns each particle a unique tag indicating whether it originated from the core or the corona. This tag is preserved in the final output and is used in our analysis to categorize the origin of the particles.

To perform multiplicity dependence studies, the minimum bias p$-$Pb sample is divided into several multiplicity percentile classes based on the total number of charged-particle counts in the pseudorapidity ($\eta$) range 2.8 $< \eta_{lab} <$ 5.1 (in the direction of the Pb beam), which is same as the ALICE-V0A detector coverage \cite{ALICEV0}.
For pp collisions, multiplicity classification is done based on the charged-particle multiplicity in the $\eta$-ranges 2.8 $< \eta_{lab} <$ 5.1 and -3.7 $< \eta_{lab} <$ -1.7 that corresponds to ALICE V0A and V0C detector acceptances, respectively.
Table~\ref{mult_properties} enlists the average charged-particle pseudorapidity densities ($<\frac{dN_{ch}}{d\eta}>$) within $|\eta_{lab}| <$ 0.5 for the different event multiplicity classes of EPOS4 simulated p-Pb events and ALICE. The agreement between EPOS4 and ALICE data is in general well found except at the top 0-5\% multiplicity class.

\begin{figure}[t]
	\centering 
	\includegraphics[width=0.48\textwidth, angle=0]{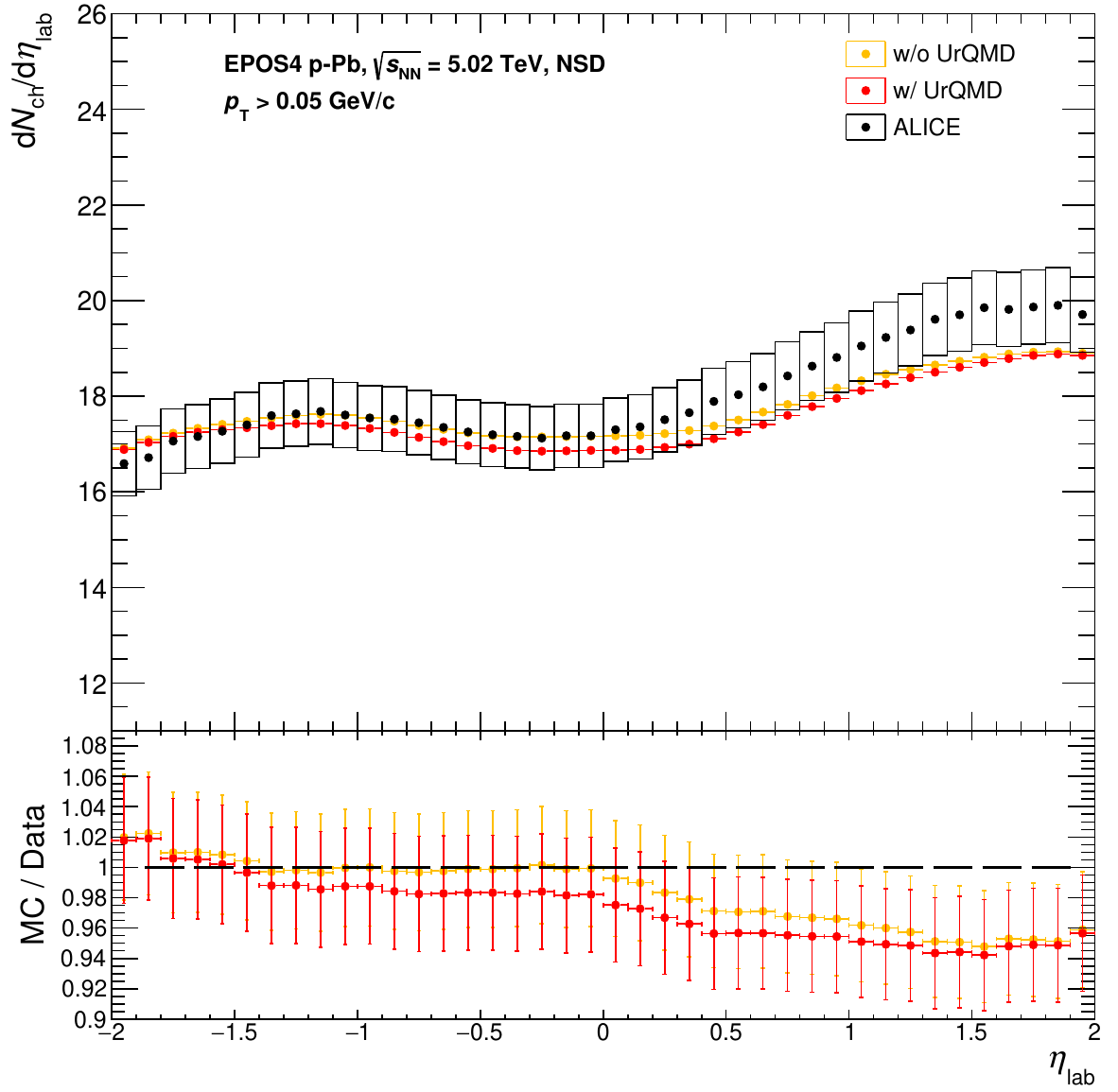}	
	\caption{(Color online)  Pseudorapidity density of charged particles in p-Pb collisions at $\sqrt{s_{\rm NN}}$ = 5.02 TeV, compared to ALICE data \cite{ALICE_pPb_etaDensity}} 
	\label{EtaDist}%
\end{figure}

To validate the generated event samples, we compare the single-inclusive transverse momentum ($p_{\rm{T}}$) spectra for the minimum bias EPOS4 simulated events with the ALICE data at mid-rapidity~\cite{ALICEPikp_pp, ALICE_piKpHighpT, Phi_ALICE}. 
Fig. \ref{comparisonptspectra} shows the $p_{\rm{T}}$-differential invariant yields of $\pi$, K, and p in pp collisions at $\sqrt{s}$ = 5.02 TeV (top) and the same for $\pi$, K, p and $\phi$ in p-Pb collisions at $\sqrt{s_{\rm NN}}$ = 5.02 TeV (bottom) from minimum bias EPOS4-simulated events, compared with ALICE results in similar kinematic ranges and detector coverages. EPOS4 is seen to describe the trend in data reasonably well but model-to-data ratios suggest a quantitative disagreement mostly at intermediate-$p_{\rm{T}}$, reaching upto 40\%. At higher-$p_{\rm{T}}$, data-to-model ratios approach unity, indicating gradual improvement in the quantitative description of the data by EPOS4 model calculation.  
We also compare the pseudo-rapidity densities of charged particles for the EPOS4-simulated p-Pb events with ALICE data, with and without hadronic re-scattering (UrQMD) effects, at $\sqrt{s_{\rm NN}}$ = 5.02 TeV in Fig. \ref{EtaDist}. The comparison suggests that the EPOS4 prediction for the mid-rapidity $\frac{dN_{ch}}{d\eta}$, with and without hadronic re-scattering effects, agrees very well even at the level 90\% or more with the experimental measurements from ALICE in the chosen kinematic interval~\cite{ALICE_pPb_etaDensity}.
Having seen that the EPOS4 calculations can emulate the $p_{\rm{T}}$- and $\eta$-differential particle production in the minimum bias pp and p$-$Pb data with reasonable quantitative accuracy in the measured kinematic ranges, we now proceed to perform the multiplicity dependence studies by classifying the minimum bias sample into several multiplicity classes as has been described above.

\section{Results}

\subsection{Transverse momentum spectra}
Transverse momentum ($p_{\rm{T}}$) spectra of produced particles are among the few basic observables that are first to be measured in any high-energy collisions. It carries essential information on the particle production mechanisms as well as on the evolutionary dynamics of the system produced in these collisions. Different physical processes that contribute to the particle production involve distinct energy scales, and therefore their contributions dominate in the different regions of the $p_{\rm{T}}$-spectra. In pp or p-A collisions, typically upto 2 GeV/c in $p_{\rm{T}}$, non-perturbative soft-QCD processes dominate and beyond 4-6 GeV/c, particles are mainly produced from the fragmentation of hard-scattered partons. The intermediate-$p_{\rm{T}}$ region, however, involves a complex and subtle interplay of both including, a new particle production based on quark recombination coming into play \cite{pTSpectra}. 
By analysing these spectra, one can therefore discern the underlying mechanisms governing particle production and explore the interplay between soft and hard processes in any high-energy collision.

\begin{figure*}[h]
    \centering
    \begin{minipage}{0.495\textwidth}
        \centering
        \includegraphics[width=\textwidth, height=0.4\textheight, angle=0]{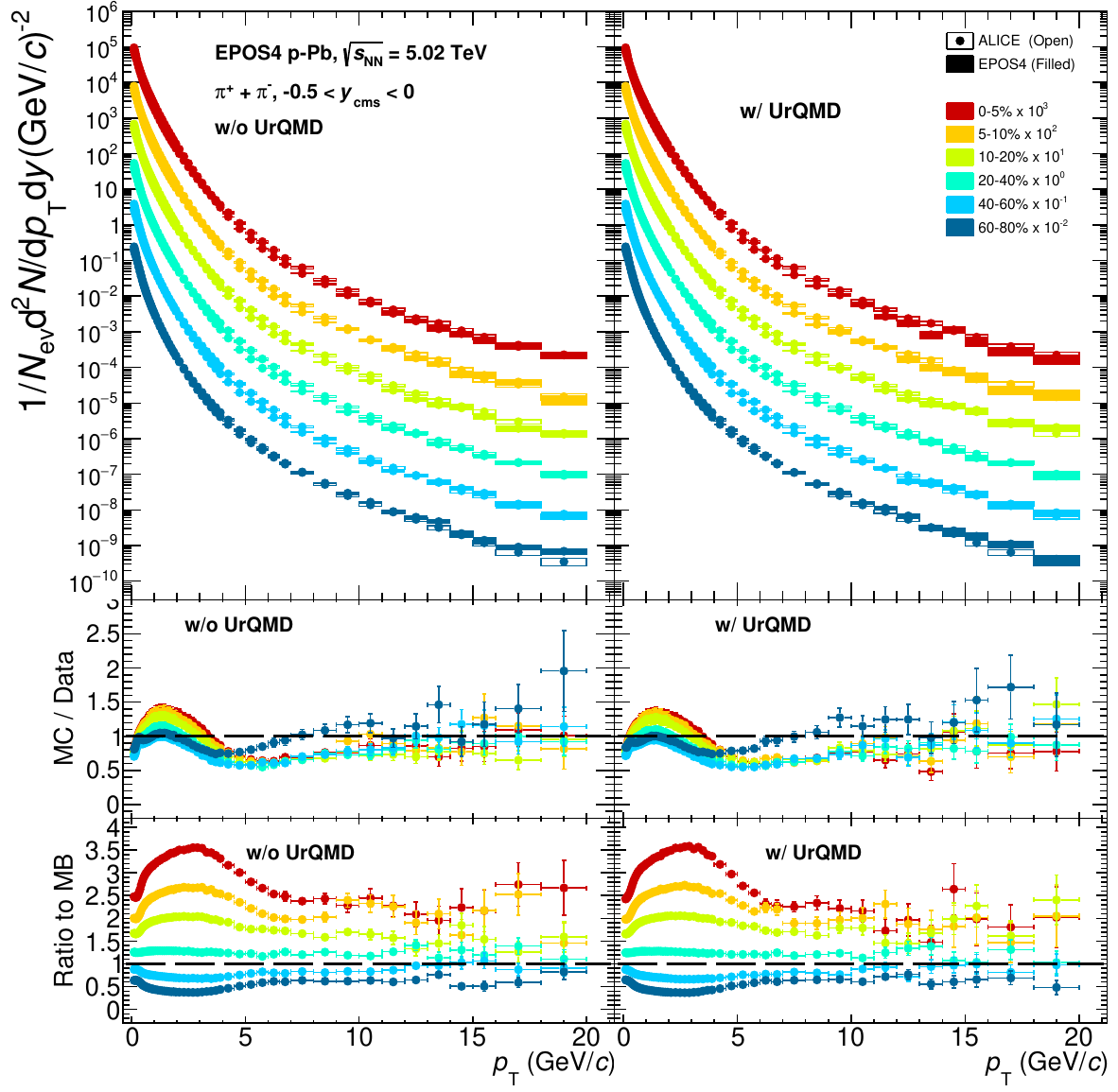}
    \end{minipage}
    \vspace{1cm} 
    \begin{minipage}{0.495\textwidth}
        \centering
        \includegraphics[width=\textwidth, height=0.4\textheight, angle=0]{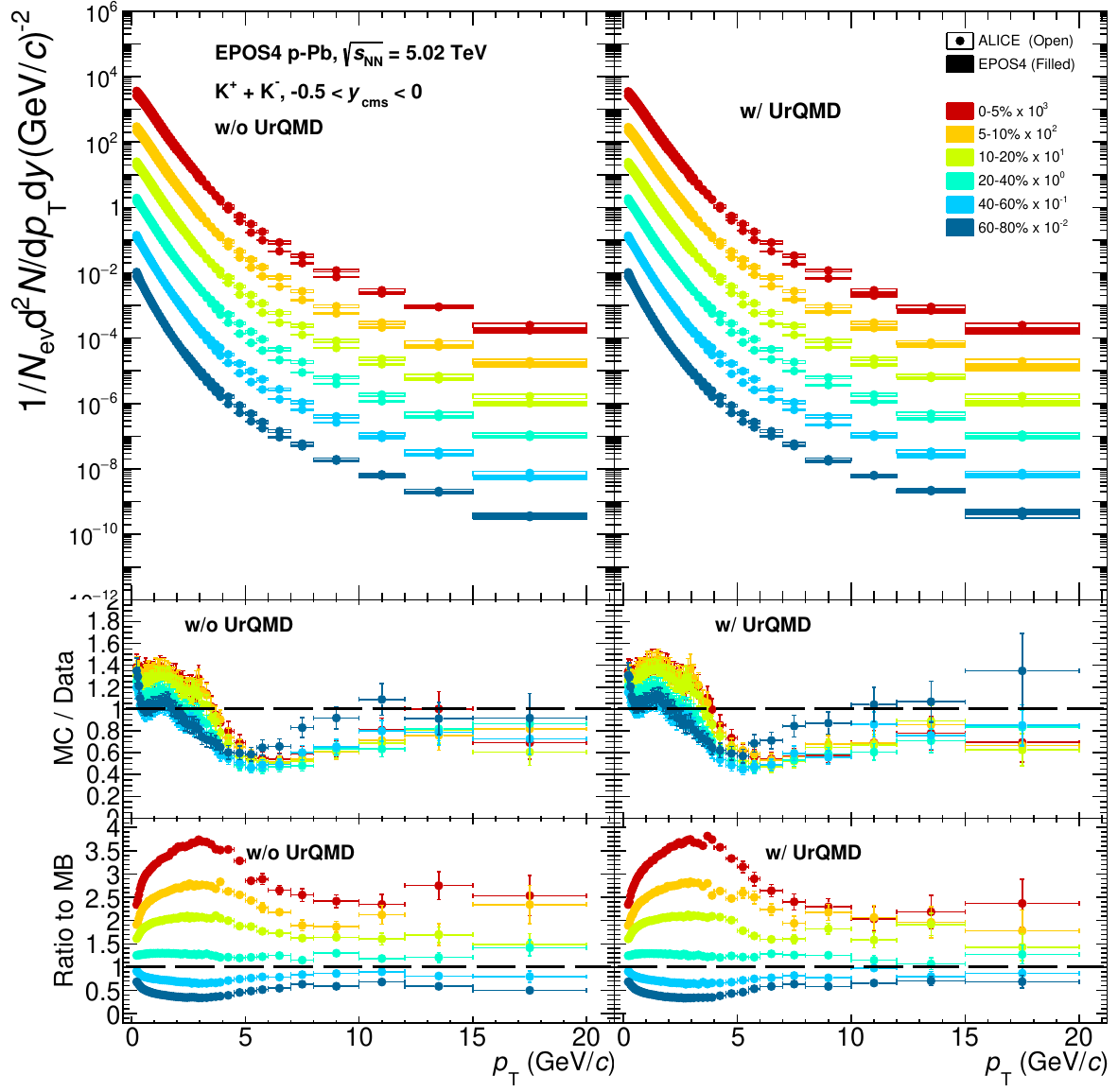}
    \end{minipage}
    \vspace{1cm} 
    \begin{minipage}{0.495\textwidth}
        \centering
        \includegraphics[width=\textwidth, height=0.4\textheight, angle=0]{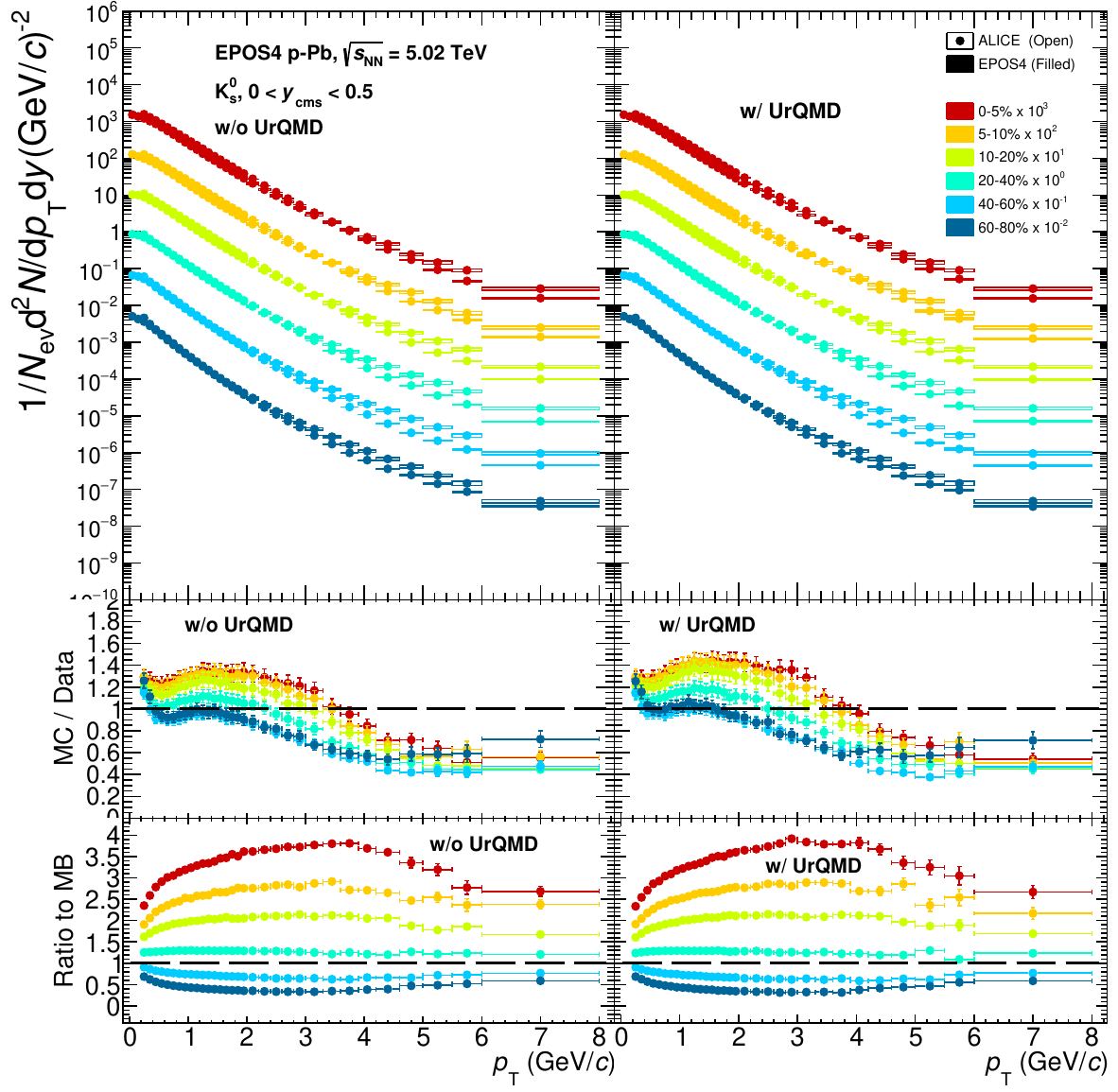}
    \end{minipage}
    \vspace{1cm} 
    \begin{minipage}{0.495\textwidth}
        \centering
        \includegraphics[width=\textwidth, height=0.4\textheight, angle=0]{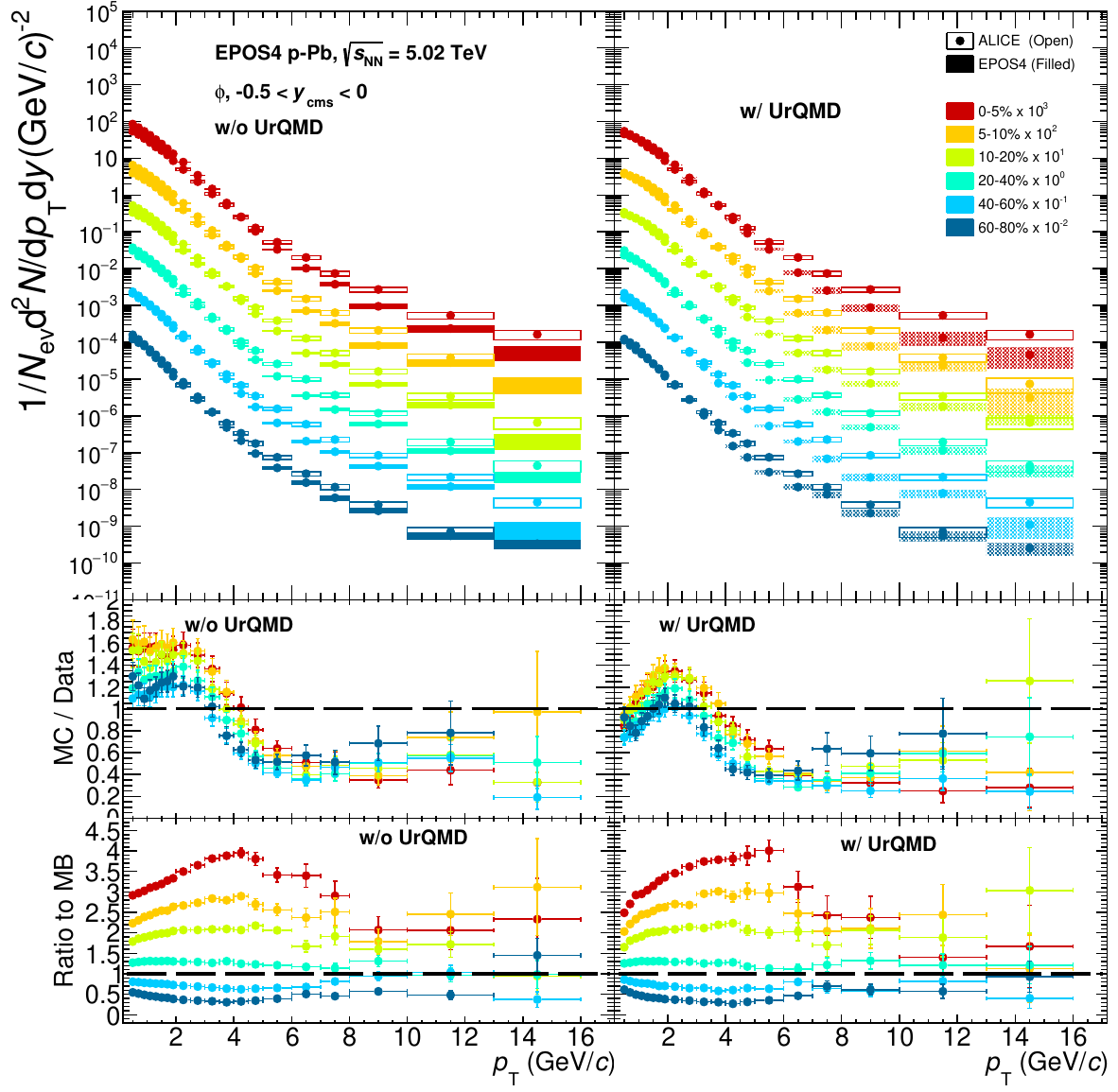}
    \end{minipage}
    \caption{(Color online) Upper: The invariant $p_{\rm{T}}$-differential spectra of mesons ($\pi$, K, $K_{s}^{0}, \phi$), compared to ALICE data \cite{ALICE_piKpHighpT, pikaprlamks0_alice, Phi_ALICE}; Middle: Ratio between EPOS4 predictions with the ALICE results; Lower: Ratio between multiplicity dependent $p_{\rm{T}}$-differential spectra to the inclusive spectra.}
    \label{pt_mesons}
\end{figure*}

\begin{figure*}[ht]
    \centering
    \begin{minipage}{0.495\textwidth}
        \centering
        \includegraphics[width=\textwidth, height=0.4\textheight, angle=0]{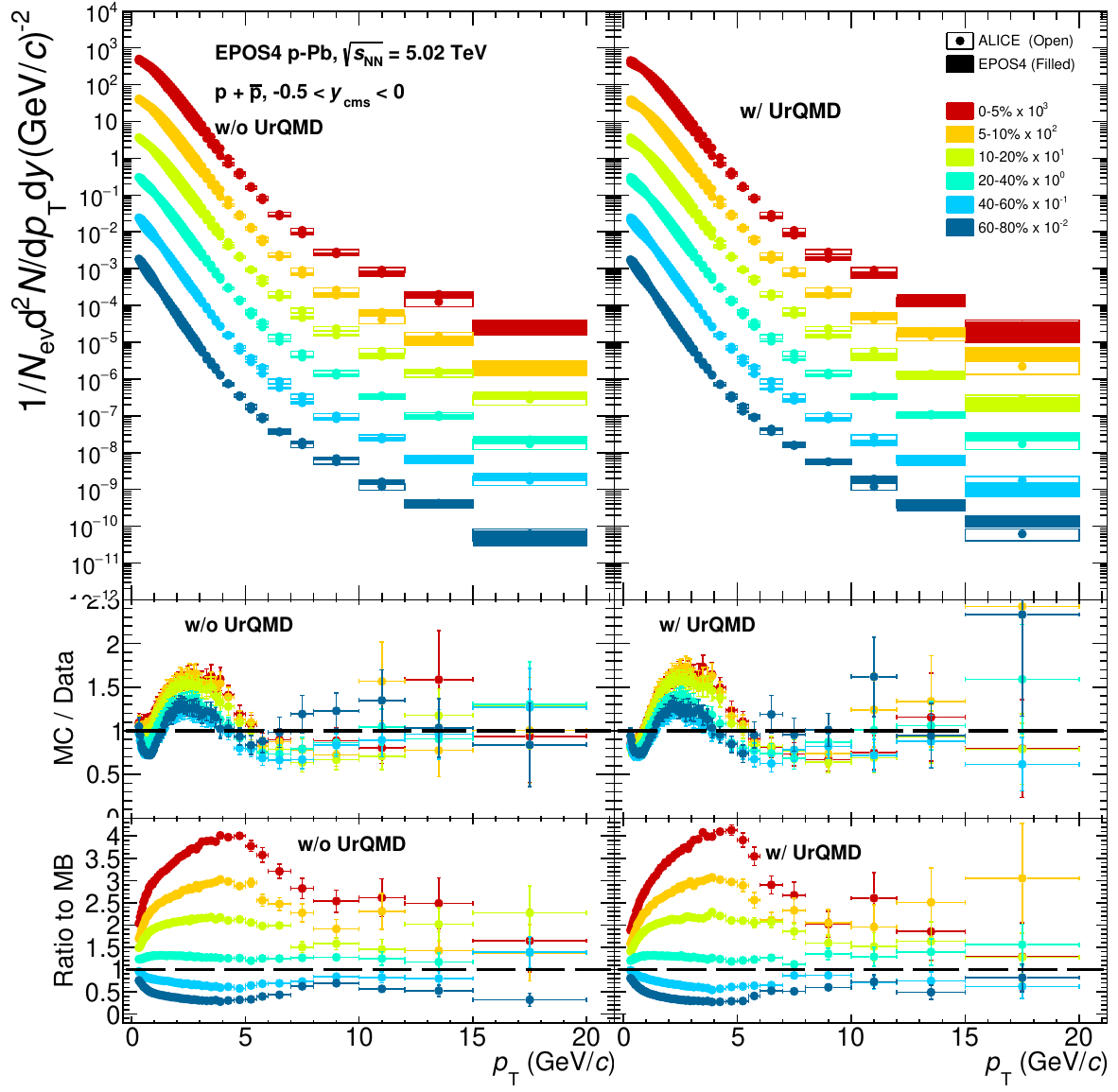}
    \end{minipage}
    \vspace{1cm} 
    \begin{minipage}{0.495\textwidth}
        \centering
        \includegraphics[width=\textwidth, height=0.4\textheight, angle=0]{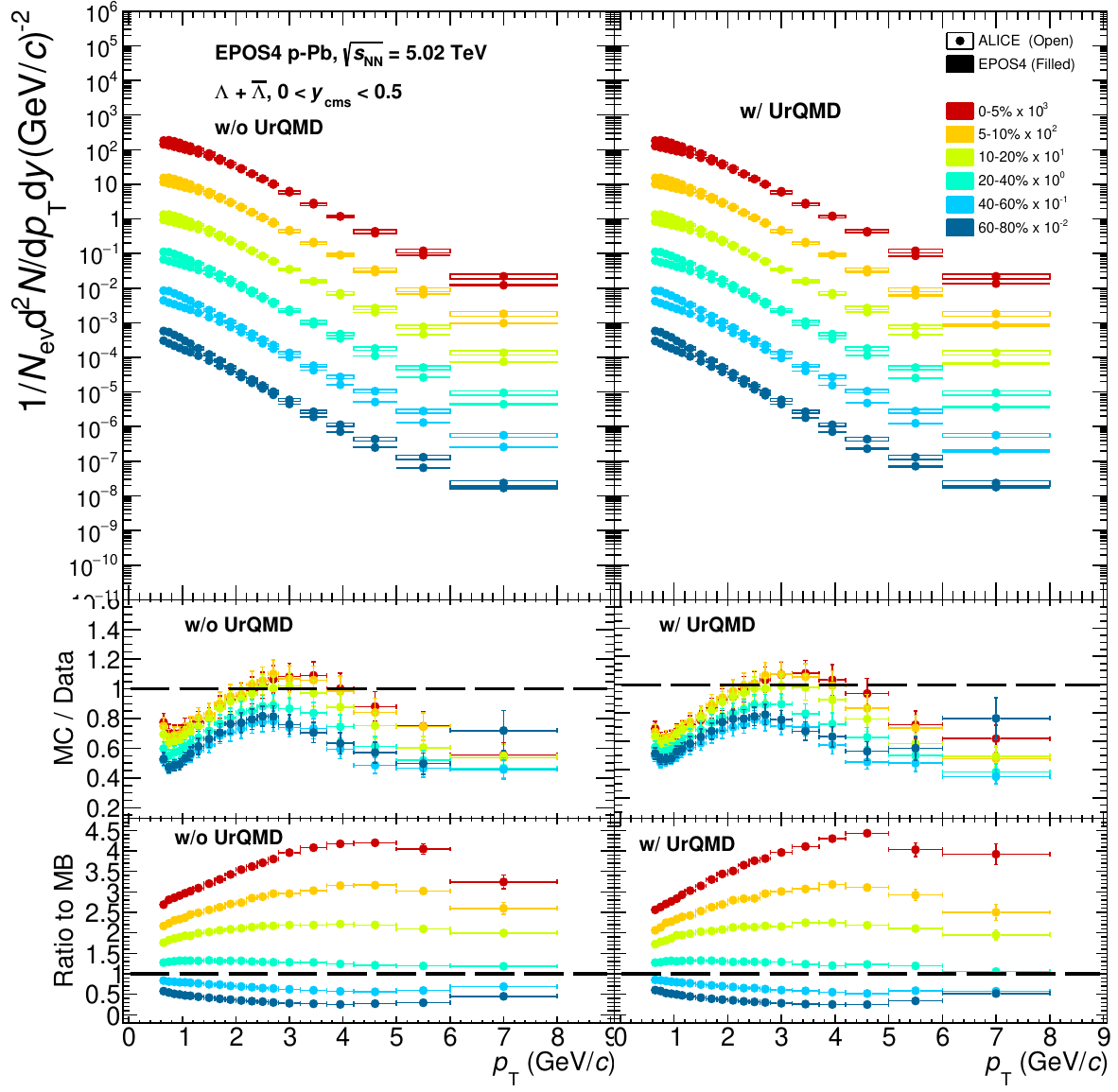}
    \end{minipage}
    \vspace{1cm} 
    \begin{minipage}{0.495\textwidth}
        \centering
        \includegraphics[width=\textwidth, height=0.4\textheight, angle=0]{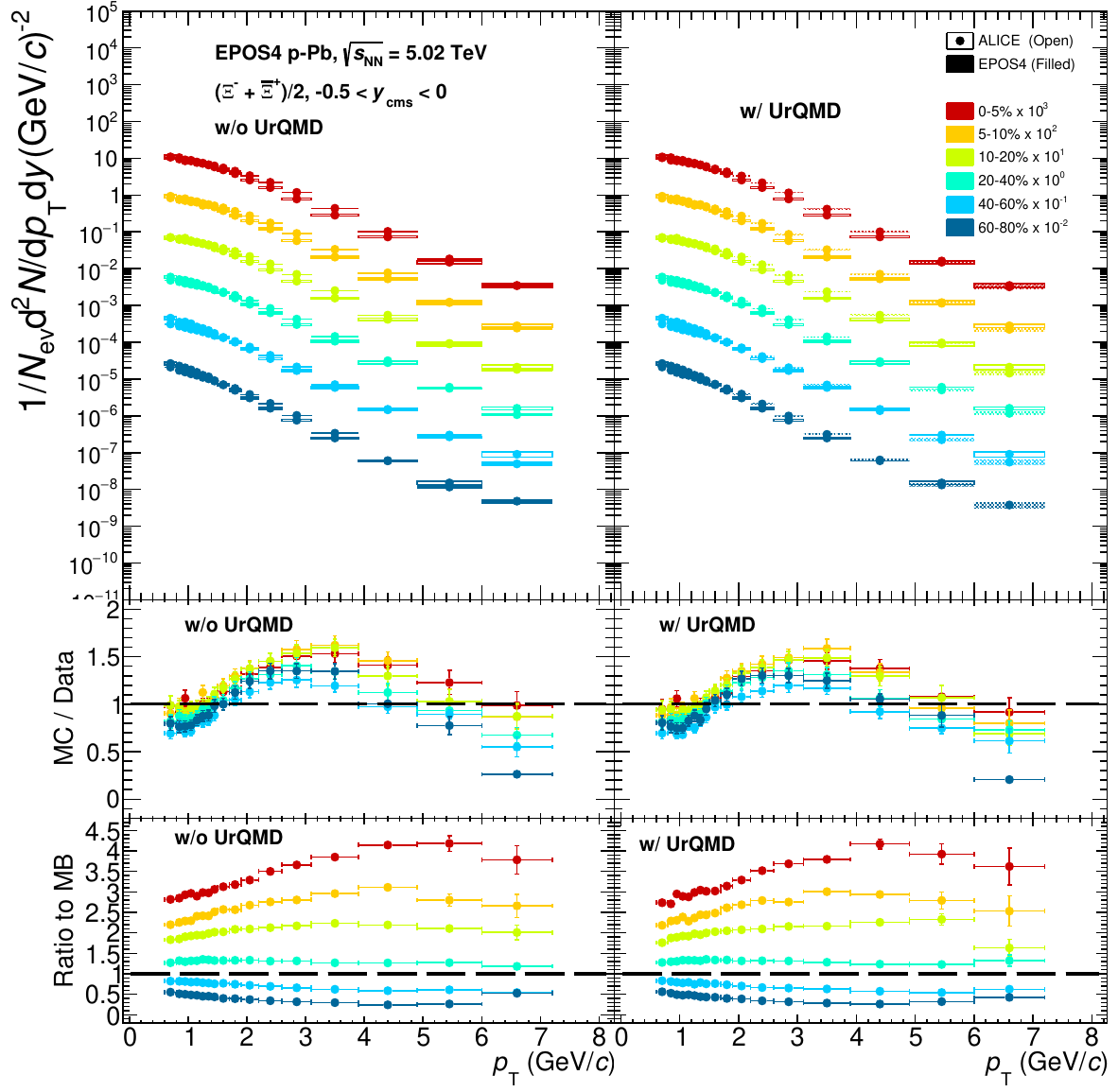}
    \end{minipage}
    \vspace{1cm} 
    \begin{minipage}{0.495\textwidth}
        \centering
        \includegraphics[width=\textwidth, height=0.4\textheight, angle=0]{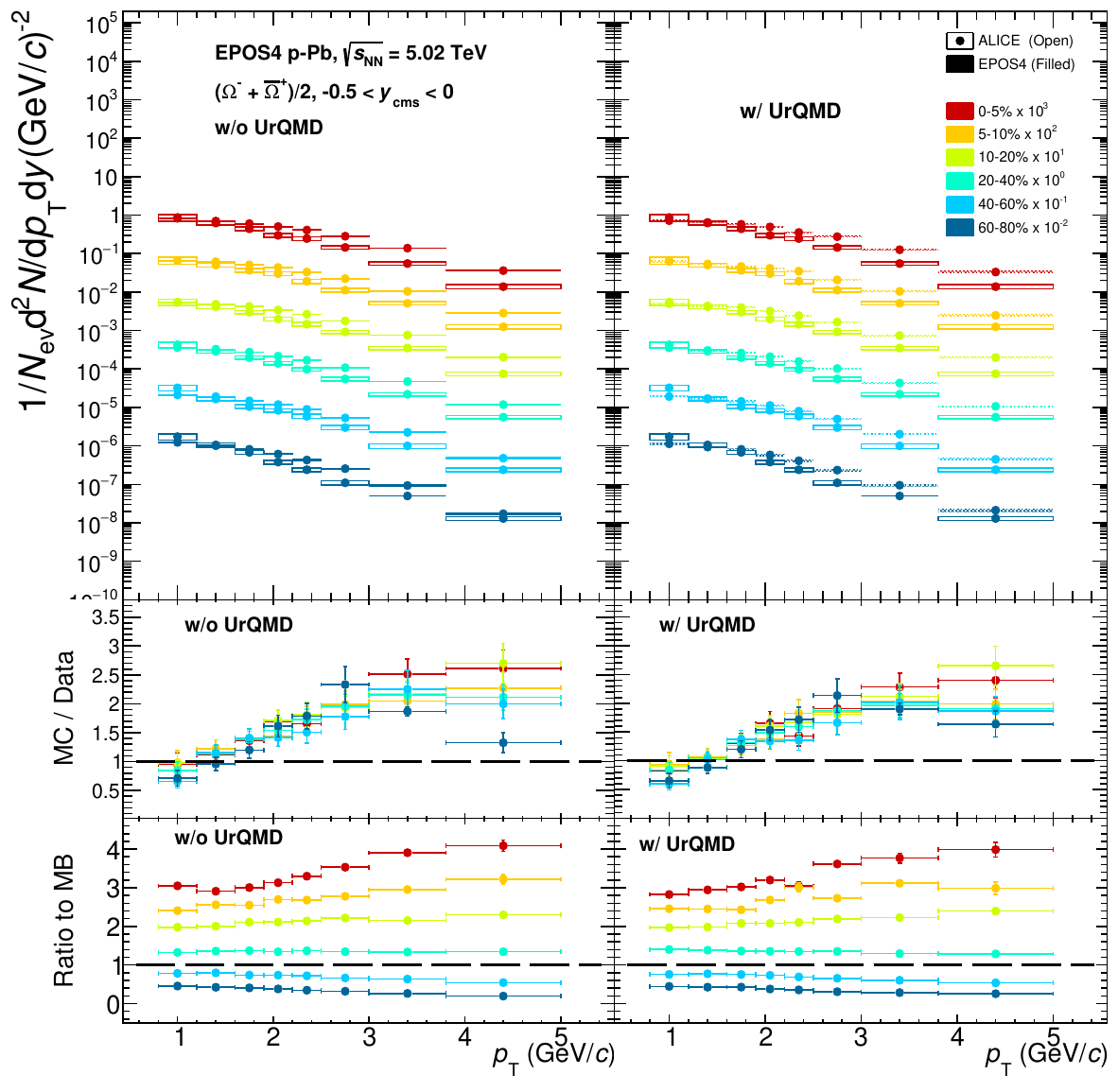}
    \end{minipage}
    \caption{(Color online) Upper: The invariant $p_{\rm{T}}$-differential spectra of baryons (p, $\Lambda$, $\Xi$, $\Omega$), compared to ALICE data \cite{ALICE_piKpHighpT, pikaprlamks0_alice, multistrange_alice}; Middle: Ratio between EPOS4 predictions with the ALICE results; Lower: Ratio between multiplicity dependent $p_{\rm{T}}$-differential spectra to the inclusive spectra.}
    \label{pt_baryons}
\end{figure*}

\begin{table*}[t]
    \centering
    \begin{tabular}{|c|c|c|c|}
        \hline
        Multiplicity Class & $\frac{dN_{ch}}{d\eta}$ (w/o UrQMD) & $\frac{dN_{ch}}{d\eta}$ (w/ UrQMD) & $\frac{dN_{ch}}{d\eta}$ (ALICE) \\ \hline
        0-5 & $50.14 \pm 0.01$ & $49.39 \pm 0.02$ & $45 \pm 1$  \\  
        5-10 & $39.71 \pm 0.01$ & $39.22 \pm 0.02$ & $36.2 \pm 0.8$ \\  
        10-20 & $31.95 \pm 0.01$ & $31.65 \pm 0.01$ & $30.5 \pm 0.7$ \\  
        20-40 & $22.42 \pm 0.01$ & $21.97 \pm 0.01$ & $23.2 \pm 0.5$ \\  
        40-60 & $14.38 \pm 0.00$ & $13.94 \pm 0.01$ & $16.1 \pm 0.4$ \\  
        60-80 & $9.64 \pm 0.00$ & $9.40 \pm 0.00$ & $9.8 \pm 0.2$ \\  \hline
    \end{tabular}
    \caption{$\frac{dN_{ch}}{d\eta}$ values within $|\eta_{lab}| <$ 0.5 for each of the six multiplicity classes  in p-Pb collisions, compared with ALICE \cite{pikaprlamks0_alice}}
    \label{mult_properties}
\end{table*}

The $p_{\rm{T}}$-distributions of $\pi$, K, p, $\Xi$, $\Omega$, $\phi$ in -0.5 $< y_{CMS} <$ 0 and $K_{s}^{0}$, $\Lambda$ in 0 $< y_{CMS} <$ 0.5 for different multiplicity intervals, according to the ALICE multiplicity classification definition for p-Pb collisions at $\sqrt{s_{\rm NN}}$ = 5.02 TeV are shown in Fig.~\ref{pt_mesons} (for mesons) and Fig.~\ref{pt_baryons} (for baryons).
In the top panels of Fig. \ref{pt_mesons} and Fig. \ref{pt_baryons}, EPOS4 predictions for the $p_{\rm{T}}$-distributions with and without hadronic re-scattering are overlaid on the ALICE results \cite{ALICE_piKpHighpT, pikaprlamks0_alice, multistrange_alice, Phi_ALICE} in the similar multiplicity classes and kinematic ranges. The right panels of Figs. \ref{pt_mesons} and \ref{pt_baryons} show the same but for EPOS4 predictions with hadronic re-scattering. To facilitate a quantitative comparison between data and EPOS4 model calculation, ratios of measured $p_{\rm{T}}$-spectra to EPOS4 predictions are shown in the middle panel of Figs.~\ref{pt_mesons}-\ref{pt_baryons}. It is seen that EPOS4 can describe the $p_{\rm{T}}$-spectra for $\pi$, K and $K_{s}^{0}$ fairly well but with a quantitative disagreement of 10-40\% at intermediate-$p_{\rm{T}}$, depending on the particle species. For $\phi$-meson agreement between data and EPOS4 is rather, not so satisfactory. Similarly for baryons, measured $p_{\rm{T}}$-spectra of p and $\Xi$ exhibit a better quantitative agreement with EPOS4 compared to $\Lambda$ and $\Omega$. EPOS4 underestimates (overestimates) the invariant $\Lambda$ ($\Omega$) yields for all multiplicity intervals over the measured $p_{\rm{T}}$-range. The origin of this disagreement may be understood, at least partially, as a limitation in modeling the relative contribution from core and corona, which will be explored further in the subsequent sections of this paper. 

To illustrate the effects of multiplicity selection on the spectral shape and $p_{\rm{T}}$-differential yields, ratios of the $p_{\rm{T}}$-spectra in a given multiplicity class to the minimum bias ones (Ratios to MB) are shown in the bottom panel of Figs.~\ref{pt_mesons}-\ref{pt_baryons}. These ratios indicate a relative enhancement or suppression of particle yields in different multiplicity classes compared to the minimum-bias sample. It is interesting to note that the $p_{\rm{T}}$-distributions exhibit a clear multiplicity dependent evolution, becoming harder with increasing multiplicity, consistently for all particle-species.

At low-$p_{\rm{T}}$, ratios of $p_{\rm{T}}$-spectra in the high multiplicity classes compared to minimum-bias ones show a significant enhancement. This enhancement is more prominent for heavier particles and at the intermediate-ranges of $p_{\rm{T}}$- a qualitative reminiscence of radial flow like effect in heavy-ion collisions, where collective expansion boosts particles with heavier mass to higher momenta\cite{MultDepSpectra, RadialFlow}. Conversely, in low-multiplicity events, these ratios show suppression, consistent with the dominance of corona contribution to particle production and consequently to the lack of significant radial flow.

The ratios of multiplicity-dependent-to-MB $p_{\rm{T}}$-differential yields in the high-$p_{\rm{T}}$ range, typically above 5-6 GeV/c, also exhibit an increasing trend from lowest to highest multiplicity class but, with a weak $p_{\rm{T}}$-dependence for all particle-species across the multiplicity classes. This indicates although, particle production changes with event multiplicity but the spectral shape, particularly in this $p_{\rm{T}}$-range, remains similar to the one in MB events. This increasing trend is also further evidence of the increasing contribution of hard processes to the high-$p_{\rm{T}}$ part of the spectra in high multiplicity event classes.

\begin{figure}
	\centering 
	\includegraphics[width=0.48\textwidth, angle=0]{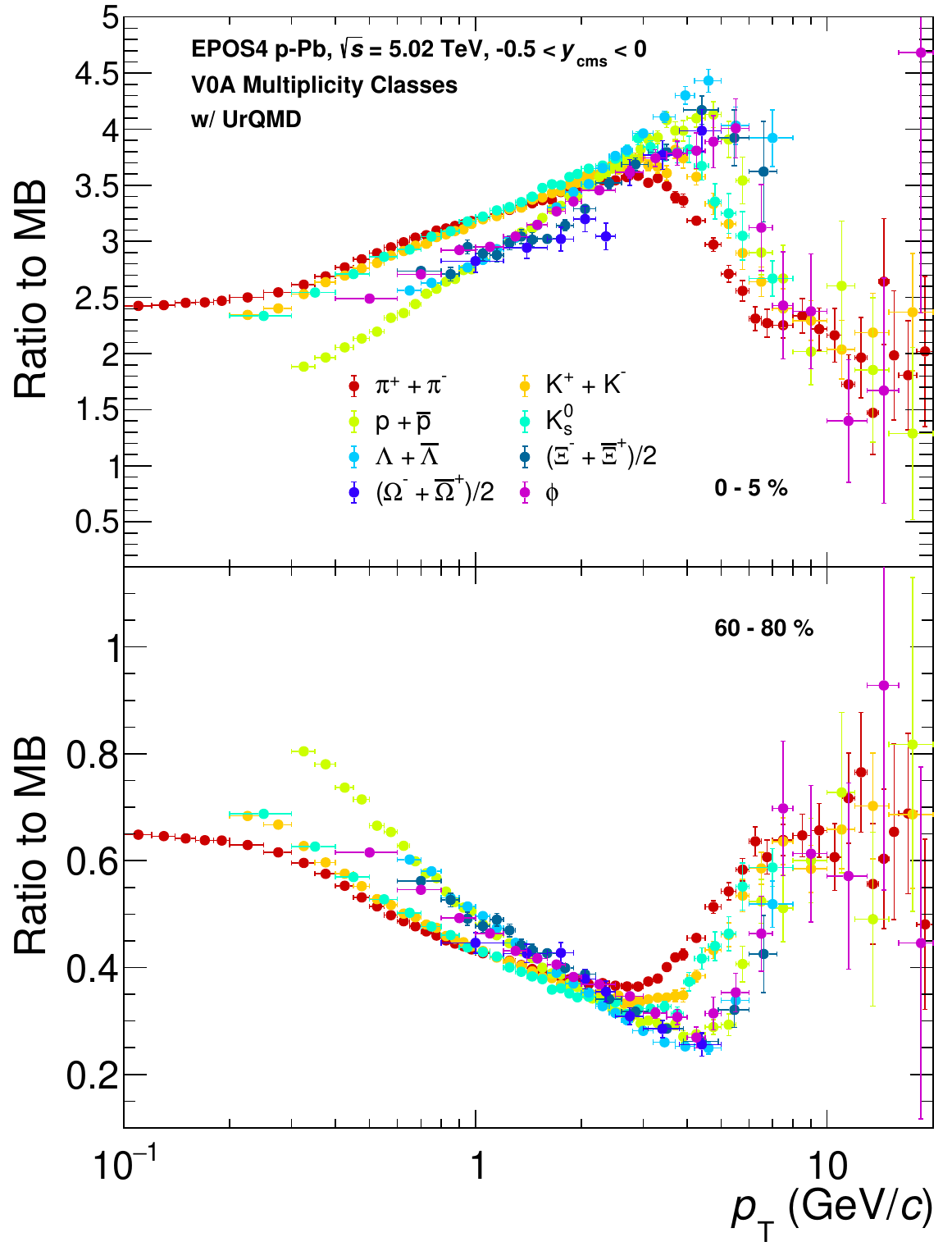}	
	\caption{(Color online) Ratio between multiplicity dependent $p_{\rm{T}}$-differential spectra to the inclusive spectra of identified hadrons for 0-5$\%$ and 60-80$\%$ multiplicity classes in p-Pb collisions} 
	\label{multRatio}%
\end{figure}

A compilation of particle species-dependent modifications of the $p_{\rm{T}}$-distribution relative to minimum bias events for 0-5\% and 60-80\% multiplicity classes from EPOS4 simulation with hadronic re-scattering is shown in Fig.~\ref{multRatio}. For all particle species, a clear hardening of the spectral shapes is observed in the high multiplicity class (0-5\%), whereas a softening occurs in the low multiplicity class (60–80\%). Interestingly, the modification of $\phi$-meson spectra follows the trend observed for multi-strange baryons rather than for mesons across the entire $p_{\rm{T}}$-range.

The observed trends in the multiplicity-dependent-to-MB ratios of $p_{\rm{T}}$-differential yields further support the role of event multiplicity in shaping the $p_{\rm{T}}$ spectra, with contributions from collective flow at low and intermediate $p_{\rm{T}}$, gradually giving way to fragmentation processes at high $p_{\rm{T}}$. Apparently, the inclusion of hadronic re-scattering effects does not significantly alter the spectral shape or the $p_{\rm{T}}$-differential yields with multiplicity, suggesting that the impact of hadronic re-scattering is nominal for small collision systems like the one produced in pp or p$-$A collisions.

\subsection{Core corona contribution to particle production for pp and p-Pb events}

\begin{figure*}[t]
	\centering 
	\includegraphics[width=0.95\textwidth, angle=0]{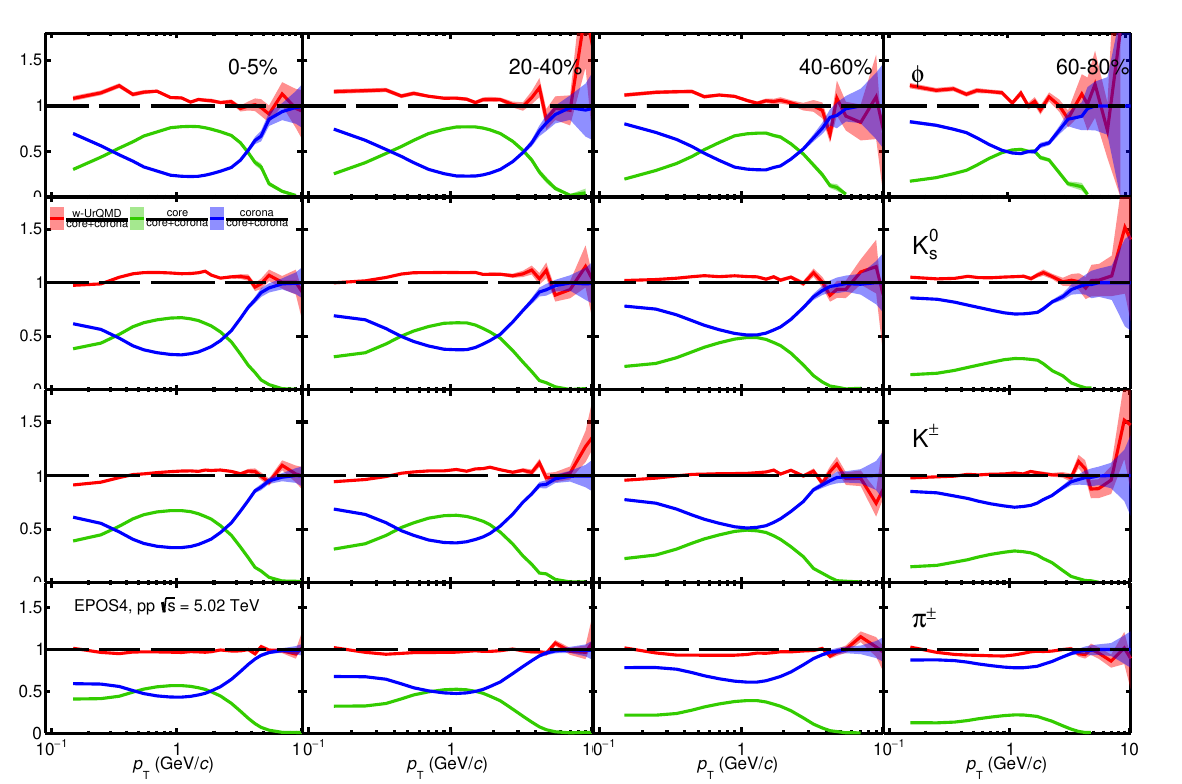}		
    \includegraphics[width=0.95\textwidth, angle=0]{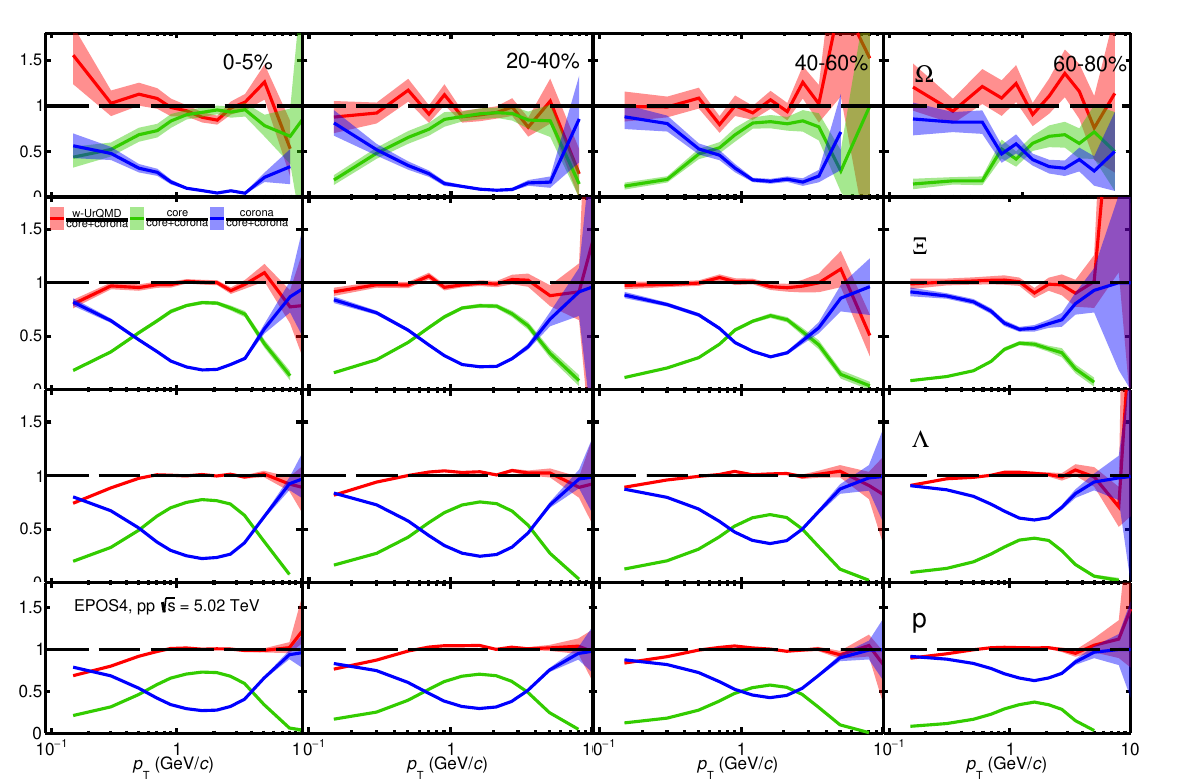}		
	\caption{(Color online) The effective ratio of X to that from w/o UrQMD, with X being the corona contribution (blue), the core (green), and the w/ UrQMD contribution (red), for four multiplicity classes, for mesons (upper panel) and baryons (lower panel), in pp collisions at $\sqrt{s}$ = 5.02 TeV.} 
	\label{core-corona-pp}%
\end{figure*}

\begin{figure*}
	\centering 
	\includegraphics[width=0.95\textwidth, angle=0]{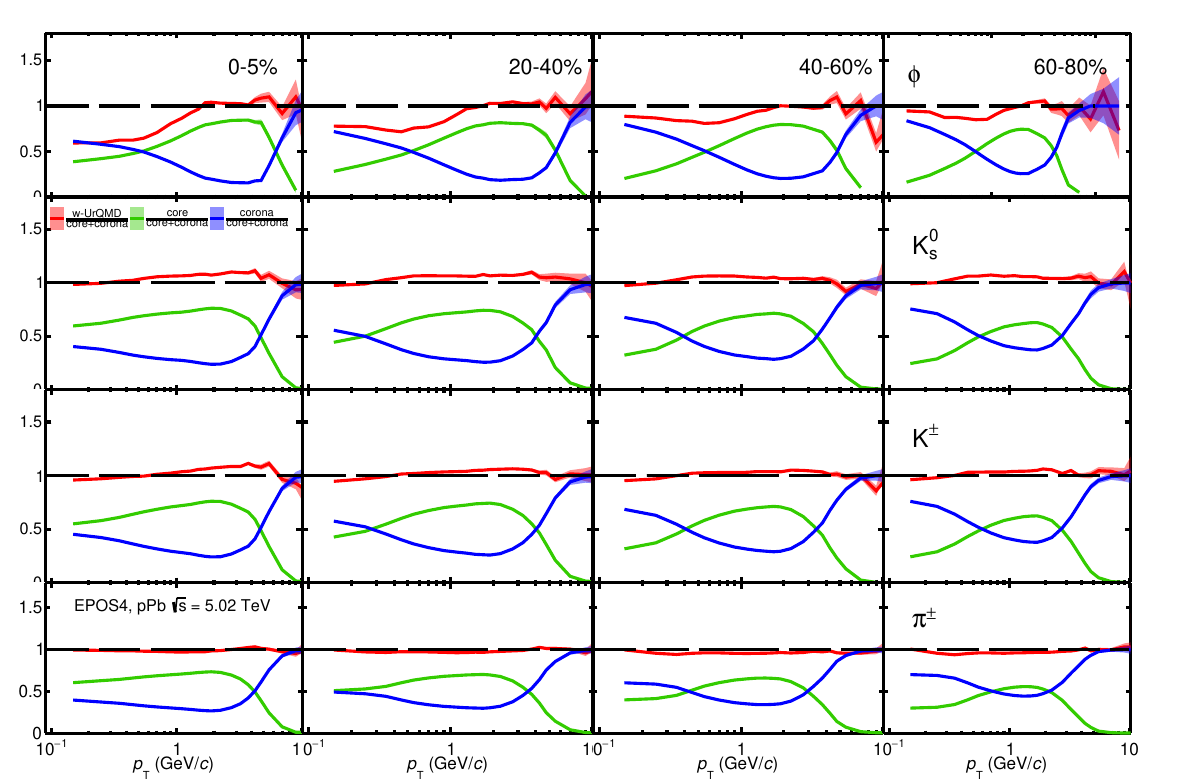}		
    \includegraphics[width=0.95\textwidth, angle=0]{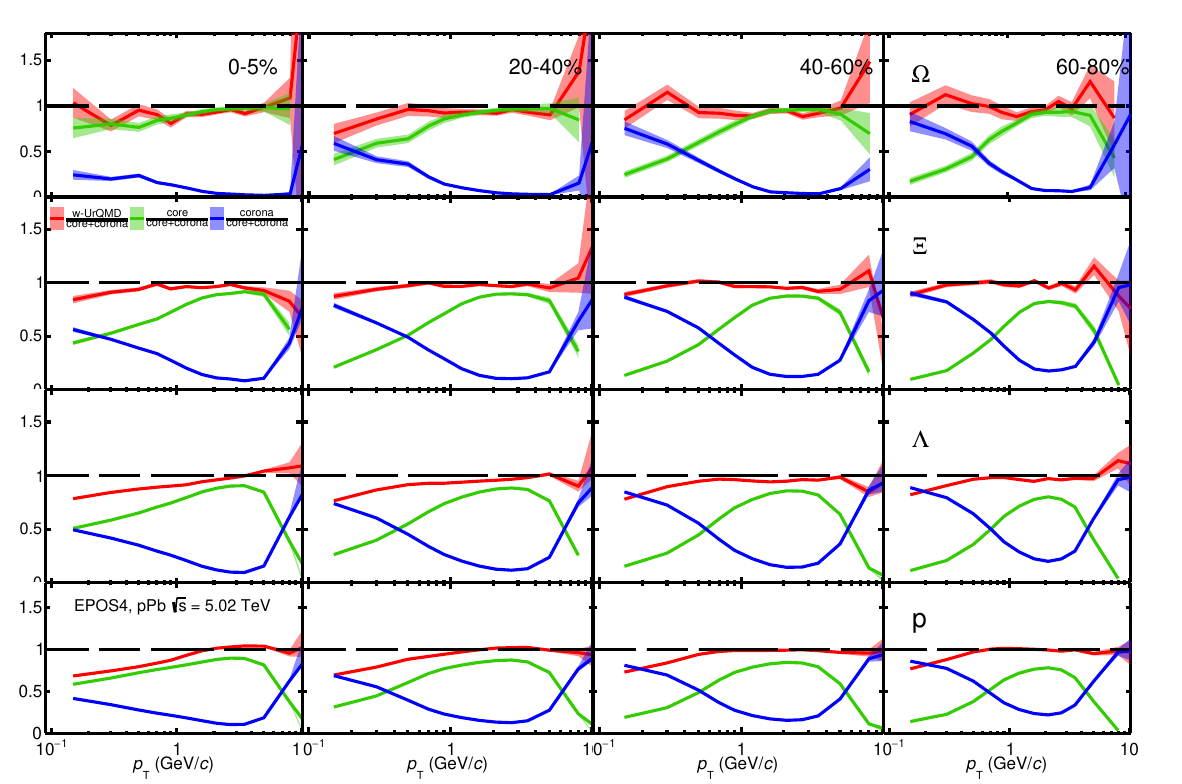}		
	\caption{(Color online) The effective ratio of X to that from w/o UrQMD, with X being the corona contribution (blue), the core (green), and the w/ UrQMD contribution (red), for four multiplicity classes, for mesons (upper panel) and baryons (lower panel), in p-Pb collisions at $\sqrt{s_{\rm NN}}$ = 5.02 TeV.} 
	\label{core-corona-p-Pb}%
\end{figure*}
As has been discussed previously, particle production in EPOS4 involves hadronization of strings invoking a special treatment of core-corona separation. Depending on the local density of string segments, a region is either classified as core or corona. The core is subjected to hydrodynamical evolution and hadronization at the freezeout hypersurface following a statistical hadronization mechanism in a microcanonical approach. Therefore, the core hadonization naturally include the effects of collective flow.  The string segments which have $p_{\rm{T}}$ sufficiently high to escape the medium without significant energy loss or those which are produced at the periphery of the medium, where the density is low, form the corona. The hadronization of the corona proceeds via the usual string fragmentation mechanism. While the core mainly contributes to the bulk of the produced particles at low $p_{\rm{T}}$, corona contribution, on the other hand, dominates at high $p_{\rm{T}}$

To understand the relative contribution of the core and corona to the overall $p_{\rm{T}}$-spectra, we calculate the $p_{\rm{T}}$-differential fractional yields for different particle species from the core and corona separately. The fractional contribution is defined as the ratio of particle yields identified to have originated from core or corona to the overall yield i.e, from both core and corona in a given $p_{\rm{T}}$ interval. 
To investigate the impact of hadronic re-scattering on the yields and the shape of $p_{\rm{T}}$ spectra, we further calculate the ratios of $p_{\rm{T}}$ spectra with and without incorporating UrQMD at the final stage. Figures \ref{core-corona-pp}-\ref{core-corona-p-Pb} show the fractional contributions from the core (green), corona (blue), and after the full evolution, i.e, taking into account the hadronic re-scattering effects (red) relative to the combined core+corona reference, as a function of $p{_T}$ for various particle-species in four multiplicity-percentile classes of pp and p-Pb collisions at $\sqrt{s_{\rm NN}} = 5.02$ TeV. 

Results from pp and p$-$Pb collisions indicate a common trend in the core-fraction (green curve), which increases from peripheral (60-80\%) to central (0-5\%) multiplicity class and, with particle mass. Besides that, both core and corona fractions exhibit a significant $p_{\rm{T}}$ and particle species dependence. In pp collisions, for all particles upto 40\% multiplicity class, relative contribution from corona dominates over core upto p$_{T}~$ 0.5-0.6 GeV/c. Subsequently, core contribution takes over and dominates until it diminishes beyond 2 to 3 GeV/c. Thereafter, corona contribution starts to increase again and beyond 3 GeV/c in $p_{\rm{T}}$, corona contributes upto 90-95\% of the overall yield. Above 40\% multiplicity percentile classes, for all mesons, excluding $\phi$, corona dominates over the core in the entire $p_{\rm{T}}$-range. For baryons, except $\Omega$, corona dominates over core throughout the full $p_{\rm{T}}$-range only in 60-80\% lowest multiplicity class. This trend is somewhat different in p$-$Pb collisions, where the core contribution is still substantial in peripheral (60-80\%) multiplicity class. 

In most of the multiplicity classes, core and corona contributions crosses each other at two points in $p_{\rm{T}}$. For low mass particles, crossing happens at lower $p_{\rm{T}}$ and vice-versa. However, in peripheral pp collisions, barring $\phi$-meson and $\Omega$-baryon, no such crossing occurs between core and corona contributions. It is also interesting to note that for $\phi$ and $\Omega$, core contribution remains significant in peripheral collisions as well. Particularly for $\Omega$, upto 70-80\% of its overall yield comes from core, reaching upto 90-95\% in the intermediate $p_{\rm{T}}$-range. This suggests that the thermal production of multi-strange particles are favoured over independent string fragmentation, where conservation of strangeness quantum number are applied at each fragmenting vertex.

In pp collisions, hadronic re-scattering effects for most particles across all multiplicity classes are found to be negligible. In p$-$Pb collision, low $p_{\rm{T}}$ yields of p, $\Lambda$ and $\phi$ are found to be suppressed towards high-multiplicity events after hadronic re-scattering are included. For p and $\Lambda$, hadronic interactions like, baryon-antibaryon annihilation is believed to be responsible for the suppression. The observed suppression of low $p_{\rm{T}}$ $\phi$-mesons is however, because of the signal loss due to lower reconstruction probability of the original resonances from their decay daughters. This happens when the decay products of resonance states scatter in hadronic medium to such an extent that it becomes impossible to reconstruct the previous resonance state. Such an effect is dominant mostly in the low $p_{\rm{T}}$ region( $<$2 GeV/c).  

It is to be mentioned that in EPOS4, to incorporate this signal-loss like effect for resonances, decay products of resonances are tracked throughout the system evolution. A resonance is declared reconstructable provided none of its decay product undergoes a re-scattering.

\begin{figure}[ht]
	\centering 
	\includegraphics[width=0.48\textwidth, height=0.5\textheight, angle=0]{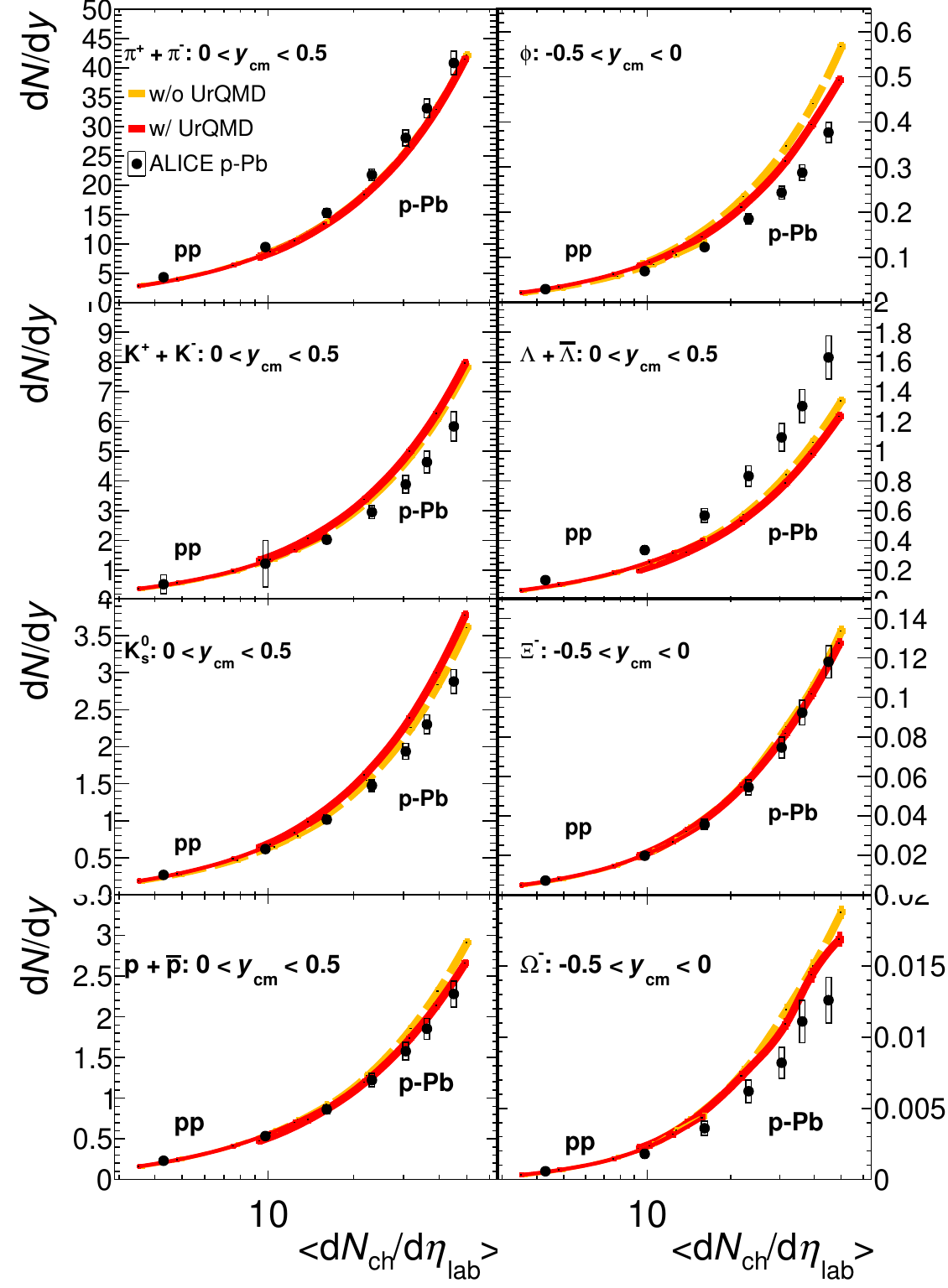}		
	\caption{(Color online) Multiplicity dependence of particle yield (dN/dy) of identified hadrons ($\pi$, K, p, $K_{s}^{0}$, $\Lambda$, $\Xi^{-}$, $\Omega^{-}$, and $\phi$) in pp and p-Pb collisions at $\sqrt{s_{\rm NN}}$ = 5.02 TeV, compared with ALICE \cite{pikaprlamks0_alice, multistrange_alice, Phi_ALICE}.} 
	\label{YieldwithNch}%
\end{figure}

\subsection{Integrated yield of identified hadrons}
Figure \ref{YieldwithNch} shows the $p_{\rm{T}}$-integrated yields ($dN/dy$) of $\pi$, K, $K_{s}^{0}$, p, and $\Lambda$ in $0 < y_{cm} < 0.5$ and $\Xi^{-}$, $\Omega^{-}$, and $\phi$ in $-0.5 < y_{cm} < 0$ as a function of average charged-particle multiplicity density ($<dN_{ch}/d\eta>$) in pp and p$-$Pb collisions at 5.02 TeV for EPOS4, with and without hadronic re-scattering. Overlaid on the same are the ALICE results \cite{pikaprlamks0_alice, multistrange_alice, Phi_ALICE} for $dN/dy$ in p$-$Pb collisions. For all particle species, a faster-than-linear increase in $dN/dy$ with $<dN_{ch}/d\eta>$, increasing smoothly from pp to p$-$Pb, is observed. EPOS4 can qualitatively reproduce this increasing trend in data but not always quantitatively. While it slightly overestimates the yields of K, $K_{s}^{0}$, $\phi$, and $\Omega^{-}$,  underestimates the same for $\pi$ and $\Lambda$. We find that turning on hadronic scattering has no significant impact, except for a small suppression of $p_{\rm{T}}$-integrated yields of $\phi$ and $\Lambda$ because of the signal loss effect and B$\bar{\rm{B}}$-annihilation, respectively.

\begin{figure}
	\centering 
	\includegraphics[width=0.48\textwidth, height=0.5\textheight, angle=0]{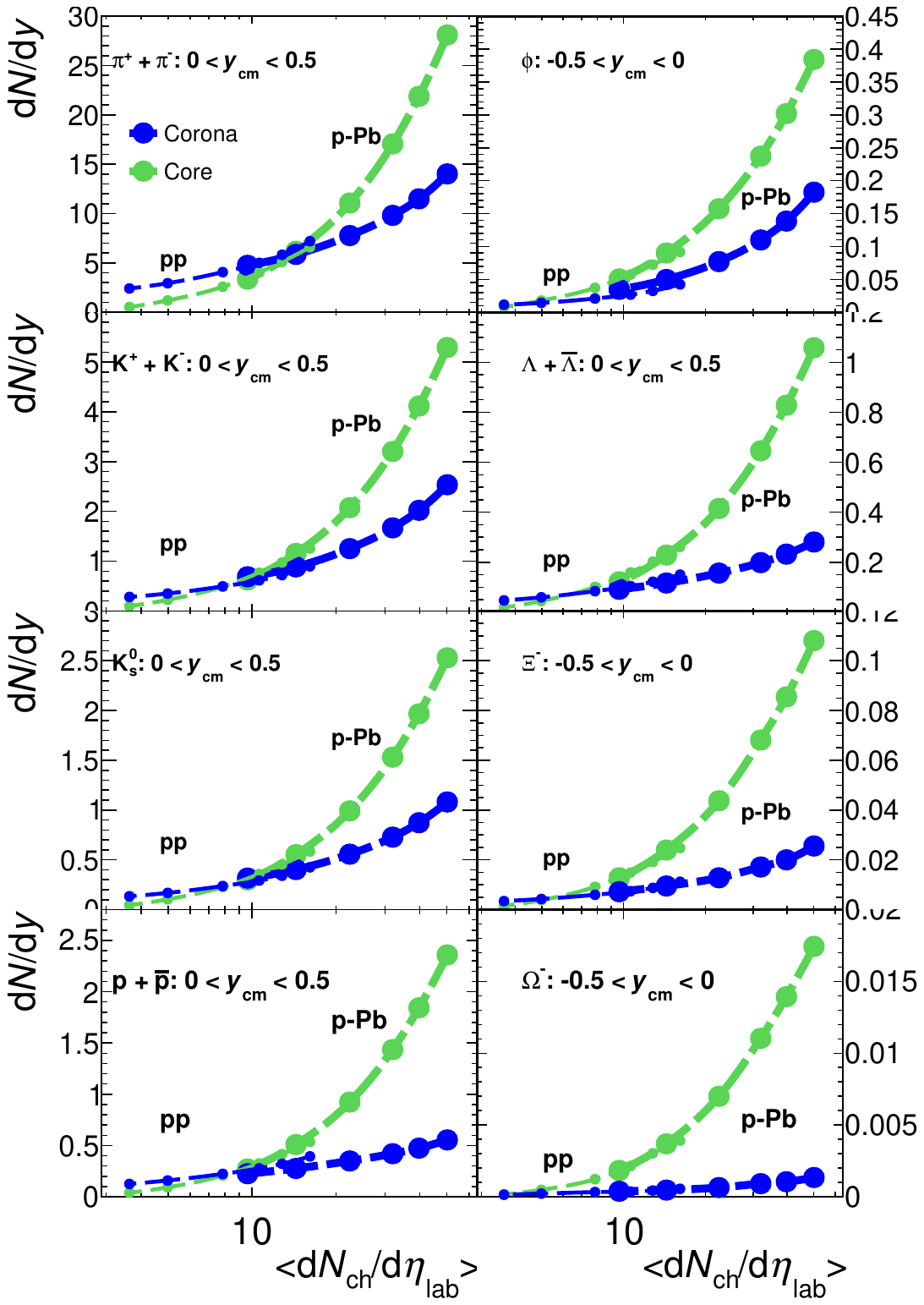}		
	\caption{(Color online) Core-corona dependence of particle yield (dN/dy) of identified hadrons ($\pi$, K, p, $K_{s}^{0}$, $\Lambda$, $\Xi^{-}$, $\Omega^{-}$, and $\phi$) in pp and p-Pb collisions at $\sqrt{s_{\rm NN}}$ = 5.02 TeV.} 
	\label{YieldwithNchwithCoreCorona}%
\end{figure}

\begin{figure}
	\centering 
	\includegraphics[width=0.48\textwidth, angle=0]{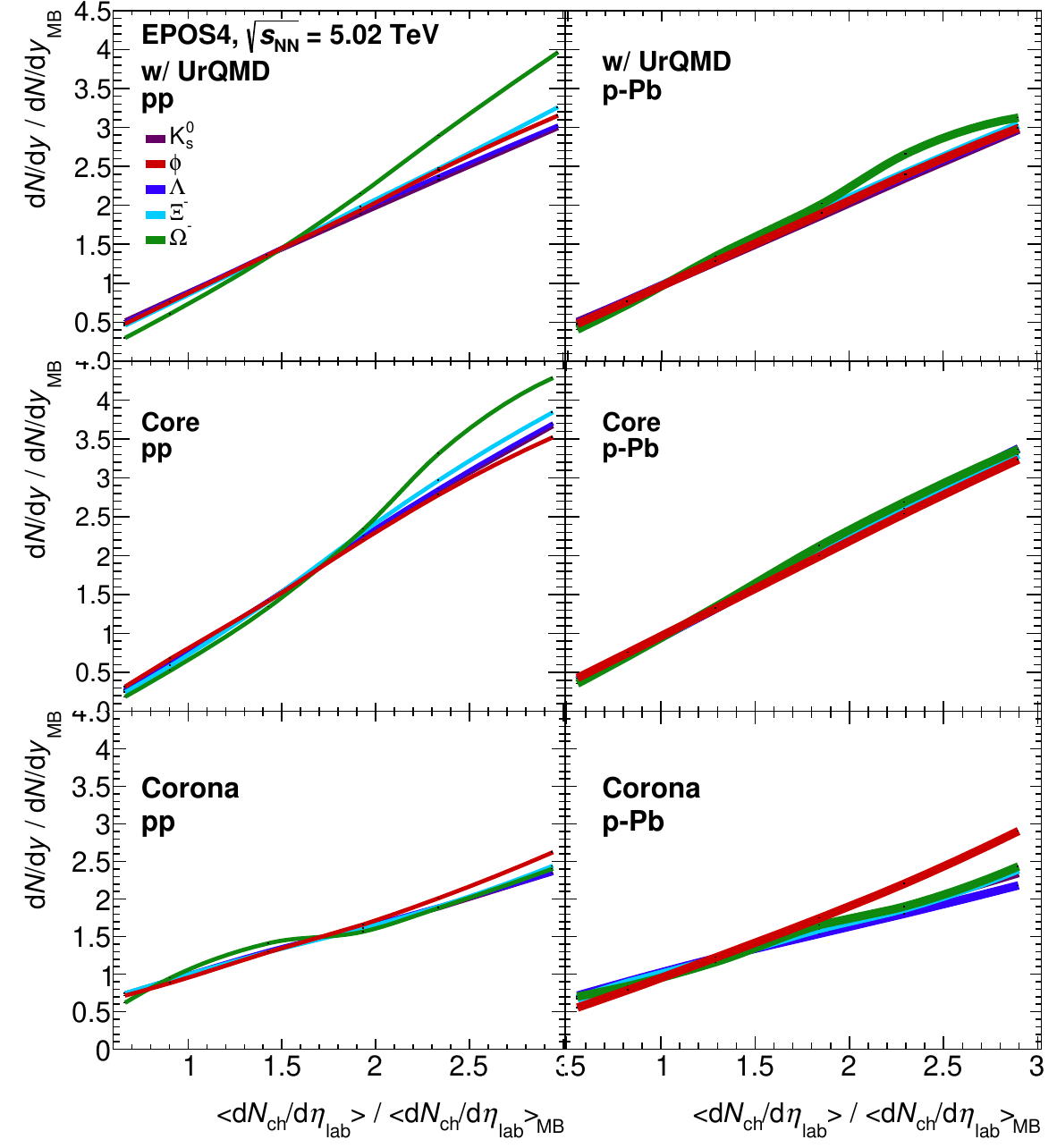}		
	\caption{(Color online) Ratio of particle yields (dN/dy) to their respective multiplicity-integrated values for (multi-)strange hadrons in pp and p-Pb collisions at $\sqrt{s_{\rm NN}}$ = 5.02 TeV, as a function of the ratio of the average charged-particle multiplicity in a given multiplicity class to that in minimum-bias events.} 
	\label{YieldRatioStrange}%
\end{figure}

Furthermore, core and corona contribution to the overall yield is estimated separately and shown in Fig.~\ref{YieldwithNchwithCoreCorona}. Both core and corona contributions to dN/dy exhibit a non-linear rise as a function of charged-particle multiplicity, with increasing dominance of core components from low to high multiplicity event classes. It is seen that core hadronization favours multi-strange hadron production, contributing more than 80\% of the total yield towards high multiplicity events.

Figure~\ref{YieldRatioStrange} presents the self-normalized yields of strange baryons and mesons (including the $\phi$-meson) as a function of the self-normalized charged particle multiplicity at midrapidity. In contrast to experimental observations \cite{ALICE2019avo}, which show that strange hadron yields increase faster than charged hadrons, the inclusive self-normalized yields of strange hadrons in EPOS4 exhibit an almost linear increase with charged particle multiplicity, except for a mild non-linear increase observed for $\Omega$ in pp collisions.

\begin{figure}
	\centering 
	\includegraphics[width=0.5\textwidth, angle=0]{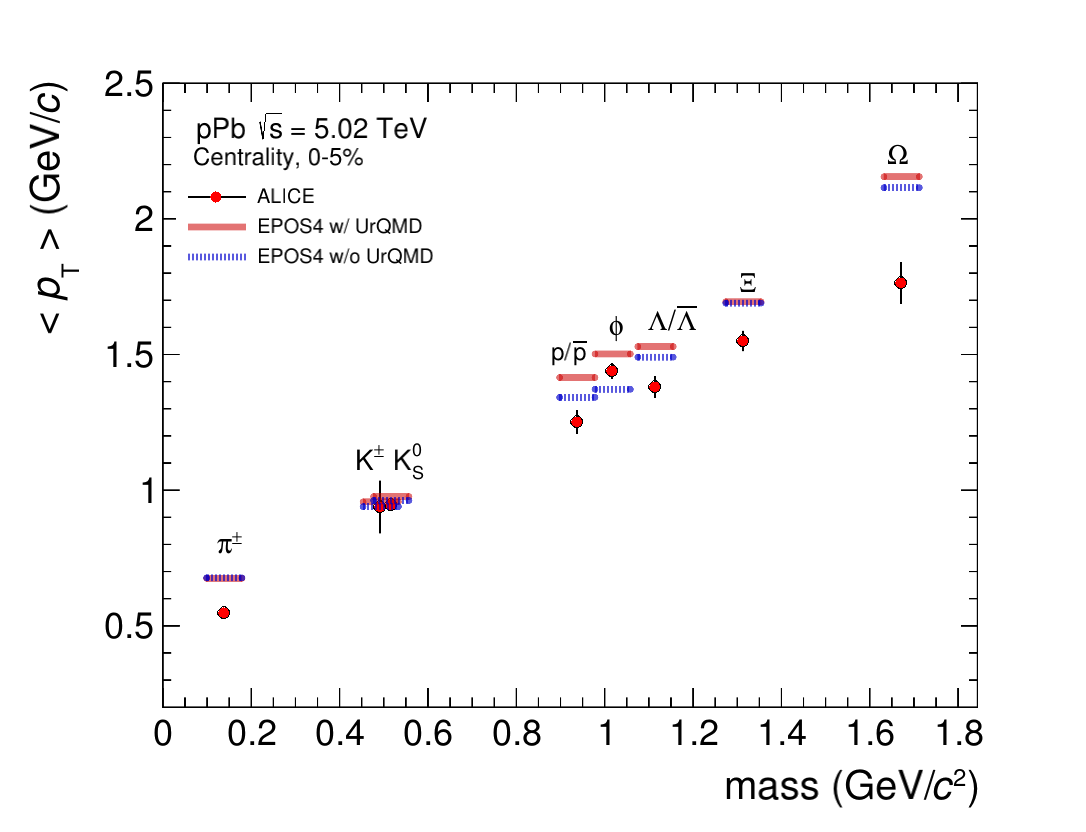}	
 \caption{(Color online) Mean transverse momentum  $\langle p_{\rm{T}} \rangle$ as a function of mass for identified hadrons for 0-5$\%$ multiplicity class as given by EPOS4 for the UrQMD OFF and ON  scenarios, compared with ALICE \cite{pikaprlamks0_alice, multistrange_alice, Phi_ALICE}.} 
	\label{meanpt-vs-mass}%
\end{figure}

\begin{figure}[h]
	\centering 
	\includegraphics[width=0.48\textwidth, angle=0]{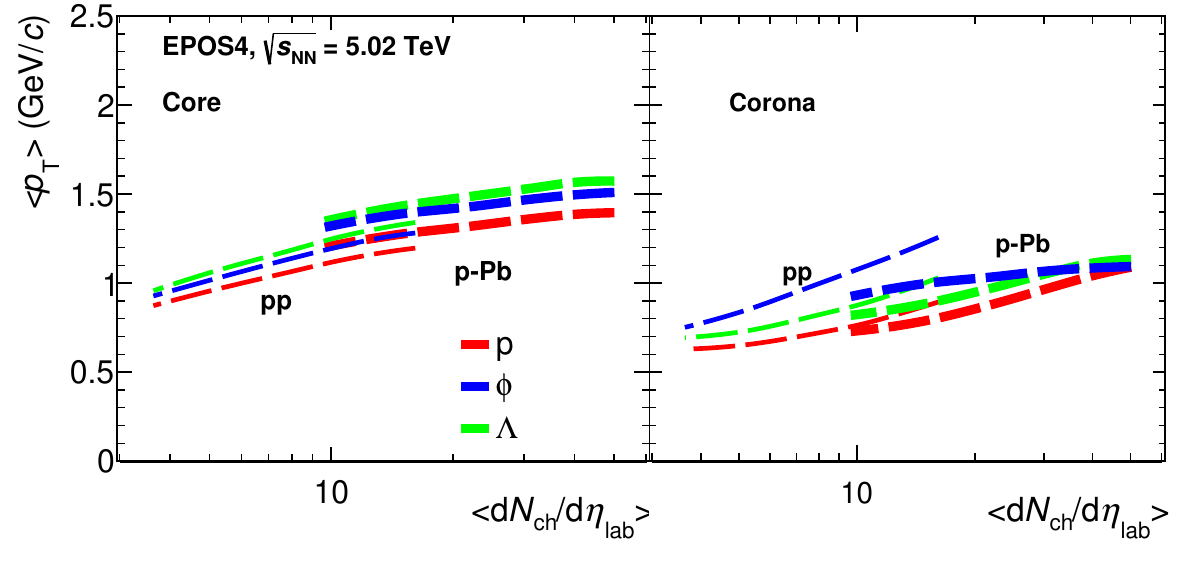}		
	\caption{(Color online) Core-corona dependence of mean transverse momentum  $\langle p_{\rm{T}} \rangle$ as a function of $\langle dN_{ch}/d\eta \rangle$ for p, $\phi$, $\Lambda$ in pp and p-Pb collisions at $\sqrt{s_{\rm NN}}$ = 5.02 TeV} 
	\label{meanptwithNch1}%
\end{figure}

\begin{figure}[h]
	\centering 
	\includegraphics[width=0.48\textwidth, height=0.5\textheight, angle=0]{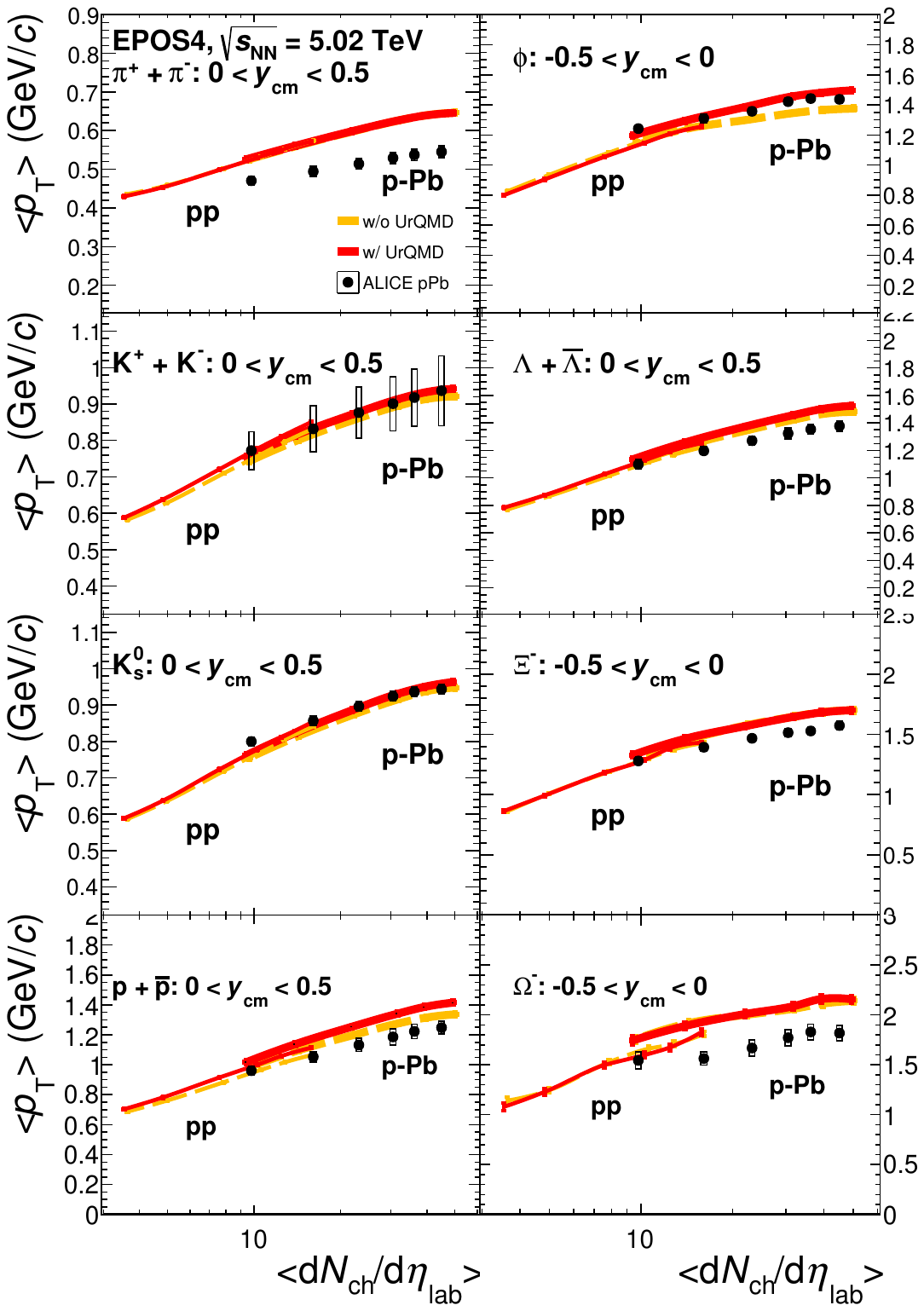}		
	\caption{(Color online) Mean transverse momentum  $\langle p_{\rm{T}} \rangle$ as a function of $\langle dN_{ch}/d\eta \rangle$ for identified hadrons given by EPOS4 for the UrQMD OFF and ON  scenarios, compared to measurements from the ALICE Experiment \cite{pikaprlamks0_alice, multistrange_alice, Phi_ALICE}} 
	\label{meanptwithNch}%
\end{figure}

\begin{figure}
	\centering 
	\includegraphics[width=0.48\textwidth, height=0.5\textheight, angle=0]{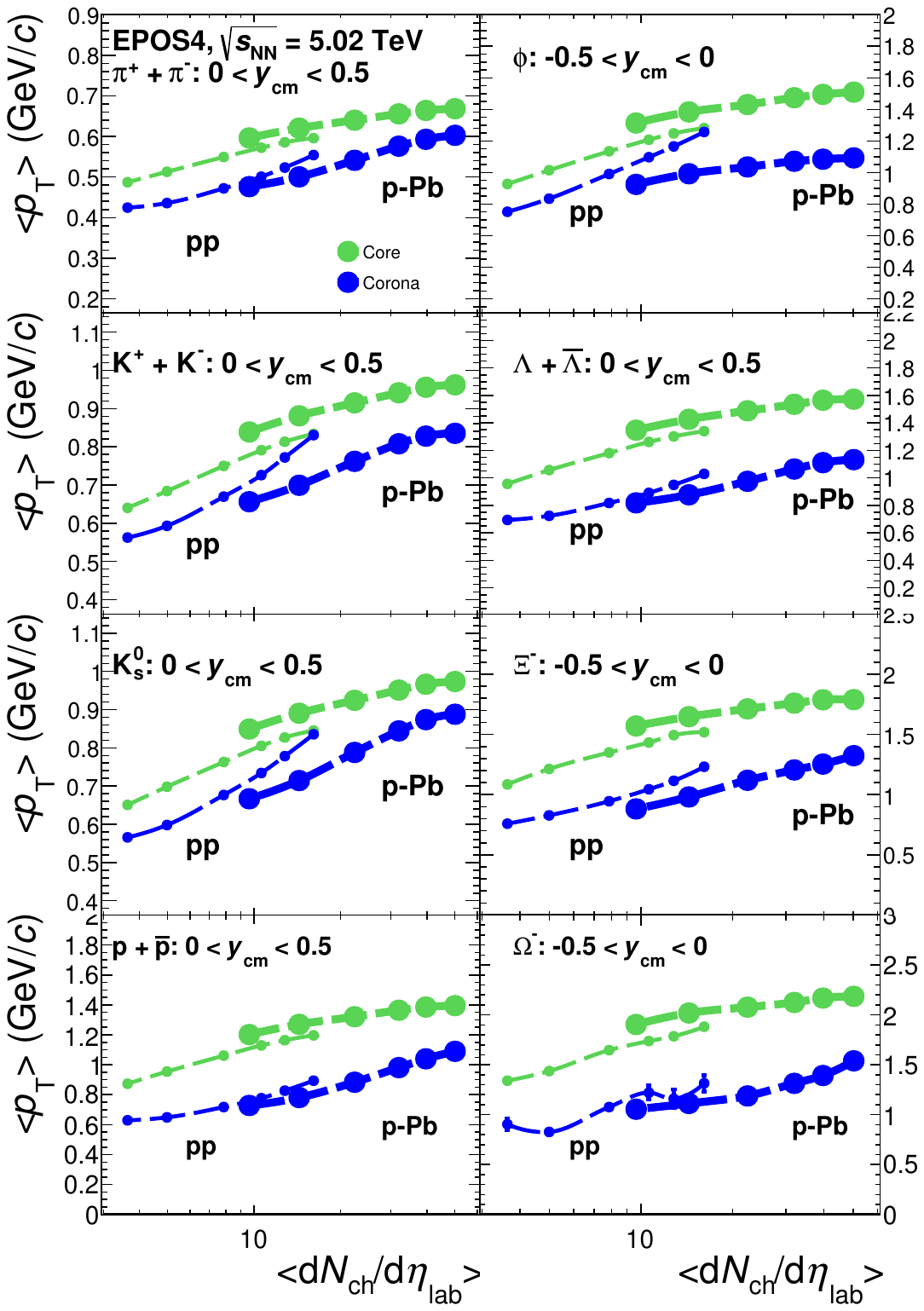}		
	\caption{(Color online) Core-corona contributions to mean transverse momentum $\langle p_{\rm{T}} \rangle$ as a function of $\langle dN_{ch}/d\eta \rangle$ for identified hadrons given by EPOS4} 
	\label{meanptwCoreCoronawithNch}%
\end{figure}

\subsection{Mean transverse momentum}
For a medium evolving hydrodynamically, the shapes of the p$_{T}-$spectra are supposedly driven by the radial expansion velocity and thus by the mass of the particle. This entails an expected 'mass ordering' in the spectral shape which is clearly seen in the measurements of average (mean) transverse momentum (\(\langle p_{\rm{T}} \rangle\)) as a function of particle mass. In other words, the mass dependence of \(\langle p_{\rm{T}} \rangle\) is a manifestation of the presence of collective (hydrodynamic) behavior in the system, although it also includes the temperature component. 
The presence of a strong radial flow is well established in heavy-ion collisions\cite{RadialFlowpp}, but for pp and p–Pb it is not yet conclusive \cite{Werner_2014}. The comparison between data and QCD-inspired models like PYTHIA rather suggests that final-state effects such as color reconnection can mimic the radial flow-like effect in pp collisions without actually assuming collectivity.

The $\langle p_{\rm{T}} \rangle$, as a function of hadron mass is calculated for various particles in high multiplicity (0–5\%) p–Pb collisions at \(\sqrt{s_{\rm NN}} = 5.02\) TeV using the EPOS4 model, both with and without UrQMD, and is shown in Fig.~\ref{meanpt-vs-mass}. The \(\langle p_{\rm{T}} \rangle\) exhibits a linear increase with particle mass, reflecting the expected mass ordering. However, the mass ordering violation observed in experimental data \cite{pikaprlamks0_alice, multistrange_alice, Phi_ALICE} for proton, \(\phi\) meson, and \(\Lambda\) is absent in EPOS4. Upon separate analysis of the core and corona components for these particles, as shown in Fig.~\ref{meanptwithNch1}, it is found that while the core \(\langle p_{\rm{T}} \rangle\) follows the ideal mass ordering, a clear violation appears for \(\phi\) mesons originating from the corona. Since the core contribution dominates the overall \(\langle p_{\rm{T}} \rangle\) versus \(\langle \frac{dN_{ch}}{d\eta} \rangle\) distribution, the mass ordering violation in the combined signal remains suppressed.

The hadronic re-scattering effects on $\langle p_{\rm{T}} \rangle$ in EPOS4 is inconspicuous for most of the particle species except for proton, $\phi$ meson and $\Lambda$. For them the re-scattering effect increases the value of $\langle p_{\rm{T}} \rangle$. 
This indicates that the spectral shapes of proton, $\phi$-meson and $\Lambda$ in EPOS4 are hardened in the hadronic phase. For proton and $\Lambda$, it happens because of the B$\bar{\rm{B}}$ annihilation of low $p_{\rm{T}}$ baryons whereas, for $\phi$-meson it is interpreted as a lower reconstruction probability of low-$p_{\rm{T}}$ $\phi$-meson due to $p_{\rm{T}}$ dependent absorption of the decay daughter hadrons in EPOS4.

We further analyze the $\langle p_{\rm{T}} \rangle$ as a function of average charged-particle multiplicity ($\langle \frac{dN_{ch}}{d\eta} \rangle$), probing the collective behavior among the produced hadrons. 
Figure \ref{meanptwithNch} shows the  $\langle p_{\rm{T}} \rangle$  of identified hadrons in pp and p-Pb collisions at $\sqrt{s_{\rm NN}} $ = 5.02 TeV for EPOS4 with and without UrQMD hadronic re-scattering. 
We observe that $\langle p_{\rm{T}} \rangle$ increases smoothly from pp to p$-$Pb with $\langle \frac{dN_{ch}}{d\eta} \rangle$ and a mass dependent increase in $\langle p_{\rm{T}} \rangle$ of the produced hadrons is clearly visible, as expected in presence of a radial flow effect.   
The results of EPOS4 calculations with and without a hadronic cascade are compared to measurements from the ALICE \cite{pikaprlamks0_alice, multistrange_alice, Phi_ALICE}.
EPOS4 in general provides a good description of the measured trends of $\langle p_{\rm{T}} \rangle$ in experimental data for p-Pb collisions, but it tends to overestimate the values of $\langle p_{\rm{T}} \rangle$ for $\pi$ and most of the baryons in p-Pb collisions.
With hadronic re-scattering effects included (that is, with UrQMD turned on), a small increase in $\langle p_{\rm{T}} \rangle$ values are observed for all particles with $\phi$ being the most prominent one. This increase in $\phi$-meson $\langle p_{\rm{T}} \rangle$ is mainly because of the hardening of the $\phi$-meson spectra due to the loss in low-$p_{\rm{T}}$ $\phi$-meson yields, as discussed previously.

Figure~\ref{meanptwCoreCoronawithNch} shows the dependence of $\langle p_{\rm{T}} \rangle$ on $\langle \frac{dN_{ch}}{d\eta} \rangle$ for core and corona components individually. Both core and corona $\langle p_{\rm{T}} \rangle$ increases with multiplicity, with the core dominating over the corona in the entire range and for all particle species. Interestingly, in the multiplicity range which is common to pp and p-Pb, the former has a higher corona $\langle p_{\rm{T}} \rangle$ for all particle species. This could be because the same $\langle \frac{dN_{ch}}{d\eta} \rangle$ value may arise from entirely different dynamics — in pp, it could be from a single hard scattering whereas, in p-Pb, it could be an aggregate effect of multiple softer interactions. Therefore, in pp collisions, the relative abundance of high-$p_{\rm{T}}$ particles may be higher than p$-$Pb even if multiplicity values are the same.

\subsection{Particle ratios}
\begin{figure}[h]
	\centering 
	\includegraphics[width=0.48\textwidth, height=0.3\textheight, angle=0]{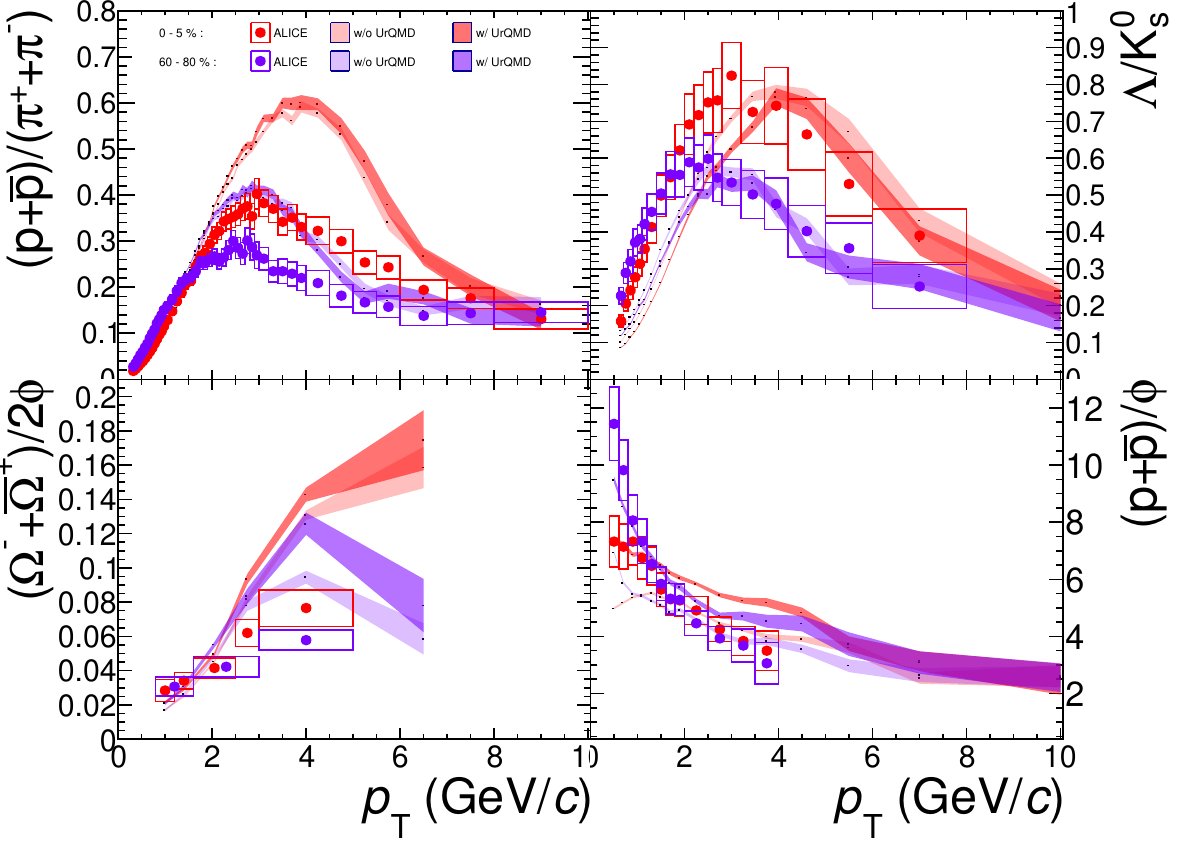}	
	\caption{(Color online) $p_{\rm{T}}$-differential baryon-to-meson ratios in p-Pb collision for 0-5$\%$ and 60-80$\%$ multiplicity class, compared with ALICE experimental results \cite{pikaprlamks0_alice, multistrange_alice, Phi_ALICE}} 
	\label{BtoMratio}%
\end{figure}

\begin{figure}
	\centering 
	\includegraphics[width=0.48\textwidth, angle=0]{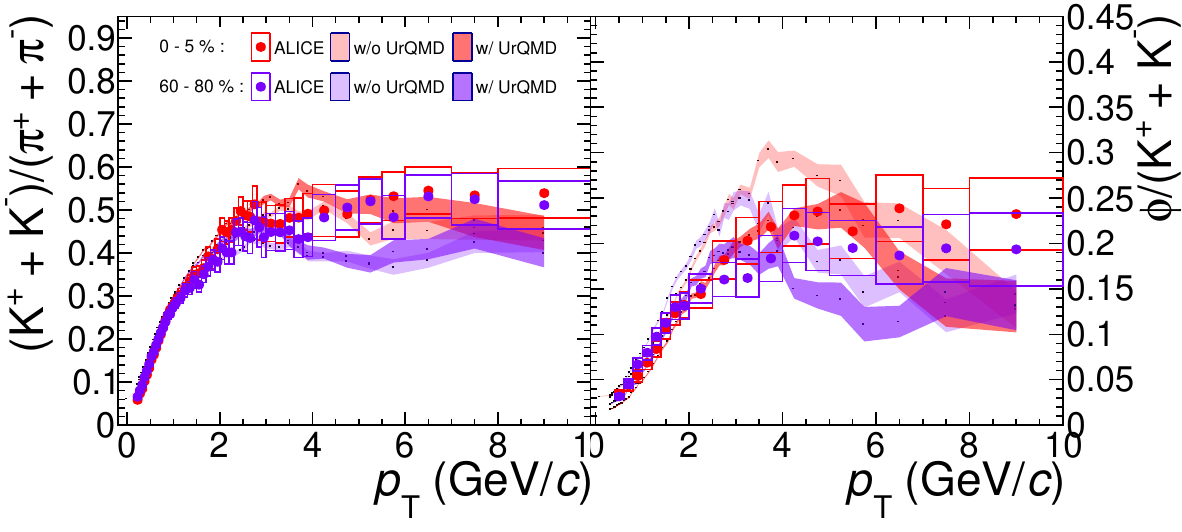}		
	\caption{(Color online) $p_{\rm{T}}$-differential K/$\pi$, $\phi$/K ratios in p-Pb collision for 0-5$\%$ and 60-80$\%$ multiplicity class, compared with ALICE experimental results \cite{ALICE_piKpHighpT, Phi_ALICE}} 
	\label{MtoMplot}%
\end{figure}

\begin{figure}[h]
	\centering 
	\includegraphics[width=0.48\textwidth, angle=0]{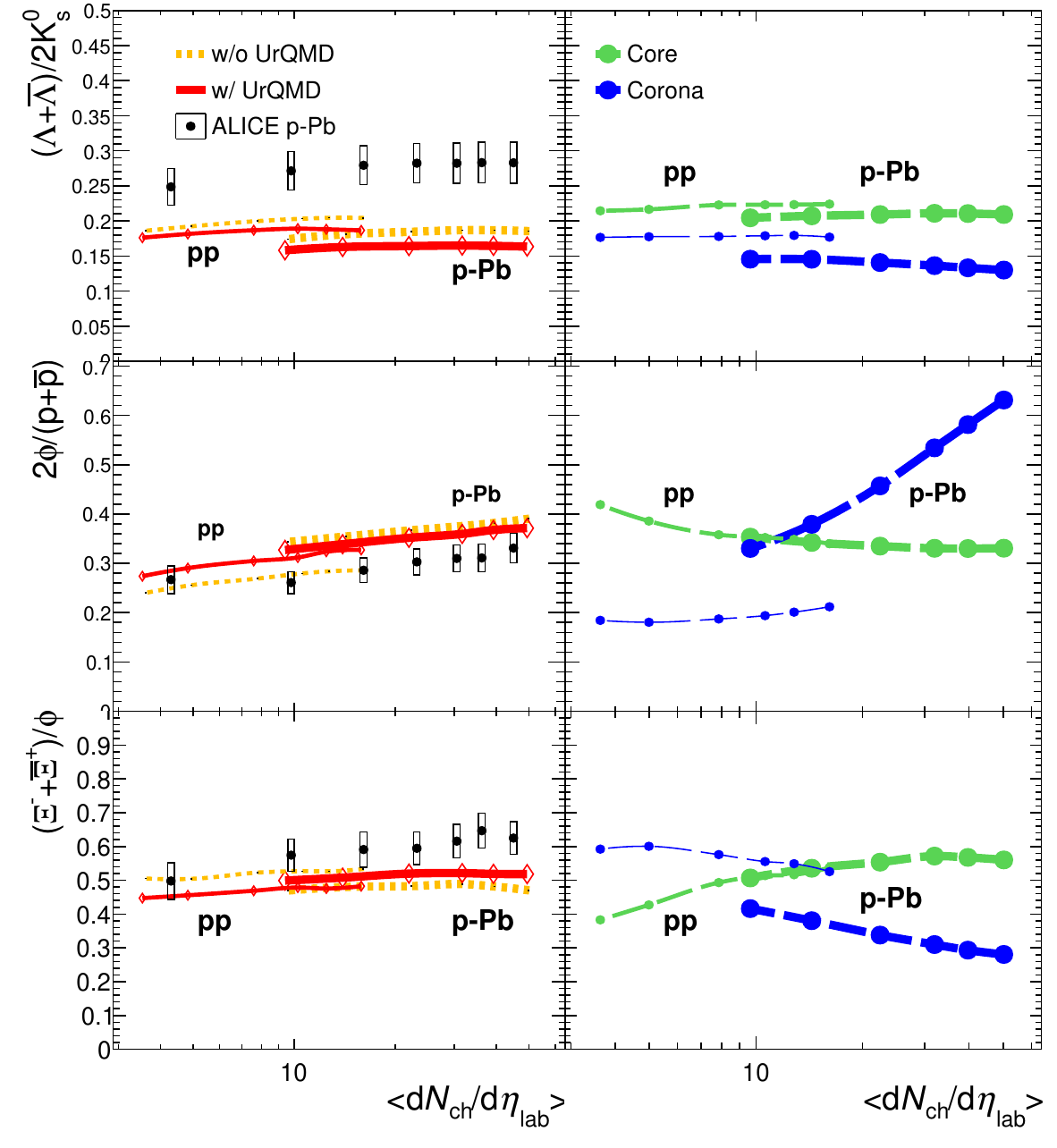}		
	\caption{(Color online) Multiplicity dependence of baryon-to-meson ratios in pp and p-Pb collision, compared with ALICE experimental results \cite{pikaprlamks0_alice, multistrange_alice, Phi_ALICE} (left) and similar ratios decomposed in core, and corona contributions (right)} 
	\label{BtoMratiowithNch}%
\end{figure}

In addition to individual particle spectra and yields, ratios of $p_{\rm{T}}$-spectra, particularly baryon-to-meson and meson-to-meson, serve as sensitive probes of the underlying particle production mechanisms. These ratios help disentangle the contributions from collective flow, quark recombination, and fragmentation processes, offering crucial insight into the dynamics of the medium produced in high-energy collisions. They also facilitate a more precise characterization of the factors that govern the final-state hadronic abundances. Furthermore, analyzing these ratios as a function of the mean charged-particle multiplicity provides valuable information on the evolution of the system’s hadrochemical composition, potentially revealing changes in the relative abundances of particle species with varying event activity.

Particle yield ratios, particularly baryon-to-meson ratios, play a crucial role in discerning whether the observed spectral shapes are primarily governed by particle mass or quark content. Such distinctions are essential for understanding the dominant particle production mechanisms, including the interplay between collective flow and hadronization processes in high-energy collisions.

Figure \ref{BtoMratio} shows a comparison of \(p_{\rm{T}}\)-differential ratios of p/$\phi$=(p+$\bar{\rm{p}}$)/$\phi$, p/$\pi$=(p+$\bar{\rm{p}}$)/($\pi^{+} + \pi^{-}$), $\Lambda/K_{s}^{0}$ and $\Omega$/$\phi$=($\Omega + \bar{\Omega}$)/2$\phi$ in 0-5$\%$ (most central) and 60-80$\%$ multiplicity classes of p$-$Pb collisions at $\sqrt{s_{\rm NN}}=$ 5.02 TeV. The results are shown for the EPOS4 model with and without the UrQMD hadronic afterburner, alongside corresponding measurements from ALICE \cite{pikaprlamks0_alice, multistrange_alice, Phi_ALICE}. In low multiplicity event classes, the p/$\phi$ ratio exhibits a decreasing trend with $p_{\rm{T}}$, whereas, in high-multiplicity events, the ratio becomes flatter, particularly at low and intermediate $p_{\rm{T}}$. Given the proton and $\phi$ meson have comparable masses, this flattening reflects the influence of radial flow in high-multiplicity event classes. Other baryon-to-meson ratios, like p/$\pi$, $\Lambda / K_{s}^{0}$ and $\Omega/\phi$ show a characteristic enhancement at intermediate $p_{\rm{T}}$, with peak position shifting towards high value in $p_{\rm{T}}$ as the multiplicity increases. Interestingly, at high $p_{\rm{T}}$, these ratios converge across multiplicity classes, suggesting a common hadronization mechanism, likely to be fragmentation, dominates in this regime. It is worth noting that while EPOS4 can reproduce the $p_{\rm{T}}$-differential $\Lambda / K_{s}^{0}$ ratios with reasonable quantitative accuracy, it significantly overestimates the p/$\pi$ ratio. This discrepancy may arise from an overestimation of the radial expansion velocity in the EPOS4 model compared to the experimental data. Since the mass difference between the proton and pion is larger than that between $\Lambda$ and K$_{s}^{0}$, the effects of radial flow are more strongly reflected in the p/$\pi$ ratio, making the deviation between data and model predictions more prominent. In the meson sector, K/$\pi$ and $\phi$/K ratios as a function of $p_{\rm{T}}$, shown in Fig.~\ref{MtoMplot}, are found to be relatively insensitive to the event multiplicity and exhibit an approximately flat $p_{\rm{T}}$ dependence above 3 GeV/c in $p_{\rm{T}}$. This flat behavior indicates that the relative changes in yields and spectral shapes between these mesons remain largely unaffected by the event activity in this momentum region despite their mass difference. Since meson production in EPOS4, in this momentum range is predominantly governed by the corona contribution, and occurs primarily through fragmentation processes, the effect of mass difference between mesons, therefore, plays a relatively minor role. Instead, the observed spectral shapes and relative yields are more strongly influenced by the quark composition of the particles. This further supports the interpretation that, at high $p_{\rm{T}}$, hadronization via fragmentation becomes the dominant mechanism, with particle abundances determined largely by parton flavor rather than collective effects such as radial flow.
This, however, is not the case for baryon-to-meson ratios, as baryon production in EPOS4 still receives a significant contribution from the core component. As a result, radial flow continues to play an important role in shaping the baryon spectra, particularly at low and intermediate $p_{\rm{T}}$. The interplay between core-driven collective effects and corona-dominated fragmentation leads to a more complex $p_{\rm{T}}$-dependence in baryon-to-meson ratios compared to meson-to-meson ratios, where fragmentation mechanisms are more dominant and less sensitive to collective dynamics. Nevertheless, at high $p_{\rm{T}}$, even the baryon-to-meson ratios exhibit an approximately flat $p_{\rm{T}}$ dependence and become largely independent of event multiplicity. This behavior indicates that, in this momentum regime, particle production is dominated by fragmentation processes which are common to both baryons and mesons, thereby minimizing the influence of collective effects.

Figure \ref{BtoMratiowithNch} shows the ratios of $p_{\rm{T}}$-integrated yields of $\Lambda/K_{S}^{0}$ and $\phi/p$ as a function of the $\langle \frac{dN_{ch}}{d\eta} \rangle$, comparing ALICE measurements \cite{pikaprlamks0_alice, multistrange_alice, Phi_ALICE} with EPOS4 predictions in pp and p-Pb collisions. The $p_{\rm{T}}$-integrated $\Lambda/K_{S}^{0}$ ratios mostly remain independent of multiplicity, indicating that the $p_{\rm{T}}$-inclusive production of $\Lambda$ and $K_{s}^{0}$ mesons follows a similar trend across different collision systems and the mass difference between the two seemingly plays no role. In contrast, there is a hint of an increase in the $\phi/p$ ratio as the multiplicity rises, implying that $\phi$-meson production may become relatively more favored compared to proton production in high-multiplicity events. These trends are further explored by examining the core and corona components separately. For $\Lambda/K_{S}^{0}$, the multiplicity dependence of both components is found to be fairly constant, whereas, for $\phi$-meson production, corona contribution within EPOS4 exhibit a significant increase, which leads to the observed enhancement in the $\phi/p$ ratio. The study of the \(\Xi/\phi\) ratio as a function of multiplicity is particularly insightful since both particles carry the same number of strange quarks. According to statistical hadronization models (SHM), the \(\Xi\) experiences canonical suppression, whereas the strangeness-neutral \(\phi\) meson does not. Therefore, the \(\Xi/\phi\) ratio is expected to increase with multiplicity. However, EPOS4 simulations show that this ratio remains approximately constant across multiplicities. Interestingly, when considering only the core component, the \(\Xi/\phi\) ratio does exhibit the predicted rise with increasing multiplicity, in agreement with SHM expectations.

\begin{figure}
	\centering 
	\includegraphics[width=0.48\textwidth, height=0.3\textheight, angle=0]{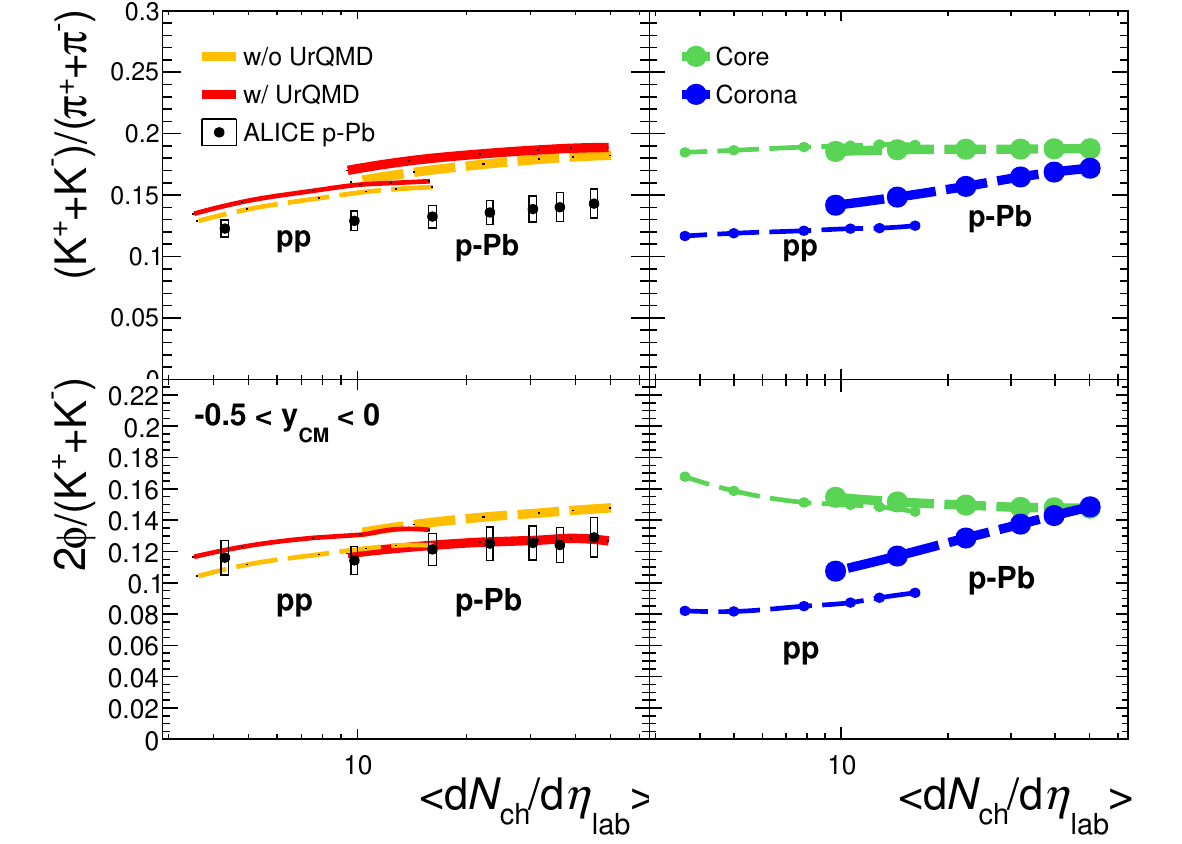}		
	\caption{(Color online) Multiplicity dependence of K/$\pi$, $\phi$/K ratios in pp and p-Pb collision, compared with ALICE experimental results \cite{ALICE_piKpHighpT, Phi_ALICE} (left) and similar ratios decomposed in the core, and corona contributions (right)} 
	\label{MtoMwithNch}%
\end{figure}

Furthermore, the meson-to-meson ratios shown in Fig~\ref{MtoMwithNch} display no significant multiplicity dependence within the uncertainties for the ALICE measurements \cite{ALICE_piKpHighpT, Phi_ALICE}. Similarly, the EPOS4 model also predicts a flat multiplicity dependence for these ratios, suggesting that the underlying processes involved in the model do not significantly alter meson production mechanisms with increasing event multiplicity, although the corona contribution to strange meson production, in particular for p-Pb collisions, as can be seen from the right panels of Fig.~\ref{MtoMwithNch} is significantly enhanced. This trend may be attributed to the fact that, with increasing event multiplicity, the average event hardness also increases—implying more energetic partonic interactions. As a result, the effective string mass in the hadronization process becomes larger, enhancing the probability of producing heavier quark combinations, such as strange quarks. This, in turn, favors the production of particles containing strange quarks, like the $\phi$-meson. In the context of EPOS4, this mechanism is particularly relevant in the corona region, where string fragmentation dominates. The increased possibility of strange quark pair production in high-mass strings could therefore explain the observed relative enhancement of $\phi$-mesons in the corona with rising multiplicity, leading to the mild increase in the overall $\phi/p$ and $\phi/K$ ratios.

\subsection{Strangeness enhancement}
\begin{figure}
	\centering 
	\includegraphics[width=0.48\textwidth, height=0.65\textheight, angle=0]{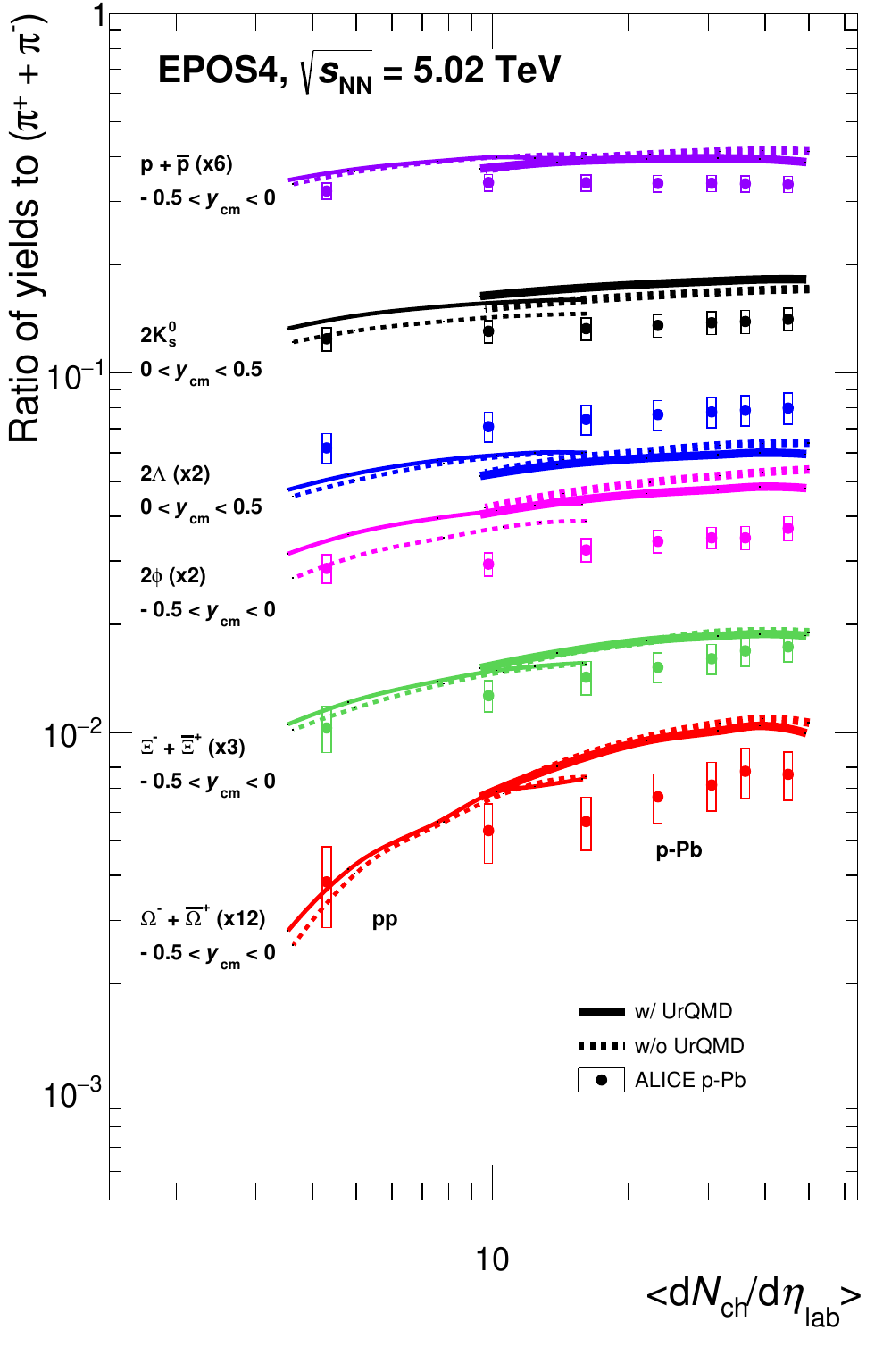}		
	\caption{(Color online) $p_{\rm{T}}$-integrated yield ratios of p, $K_{s}^{0}$, $\Lambda$, $\Xi^{-}$, $\Omega^{-}$, and $\phi$ to pions $(\pi^{-} + \pi^{+})$ as a function of $\langle \frac{dN_{ch}}{d\eta} \rangle$ in pp and p-Pb collisions at $\sqrt{s_{\rm NN}}$ = 5.02 TeV using EPOS4, The values are compared to the published results from p-Pb collisions \cite{pikaprlamks0_alice, multistrange_alice, Phi_ALICE}} 
	\label{SEwithNch}%
\end{figure}

\begin{figure}
	\centering 
	\includegraphics[width=0.48\textwidth, angle=0]{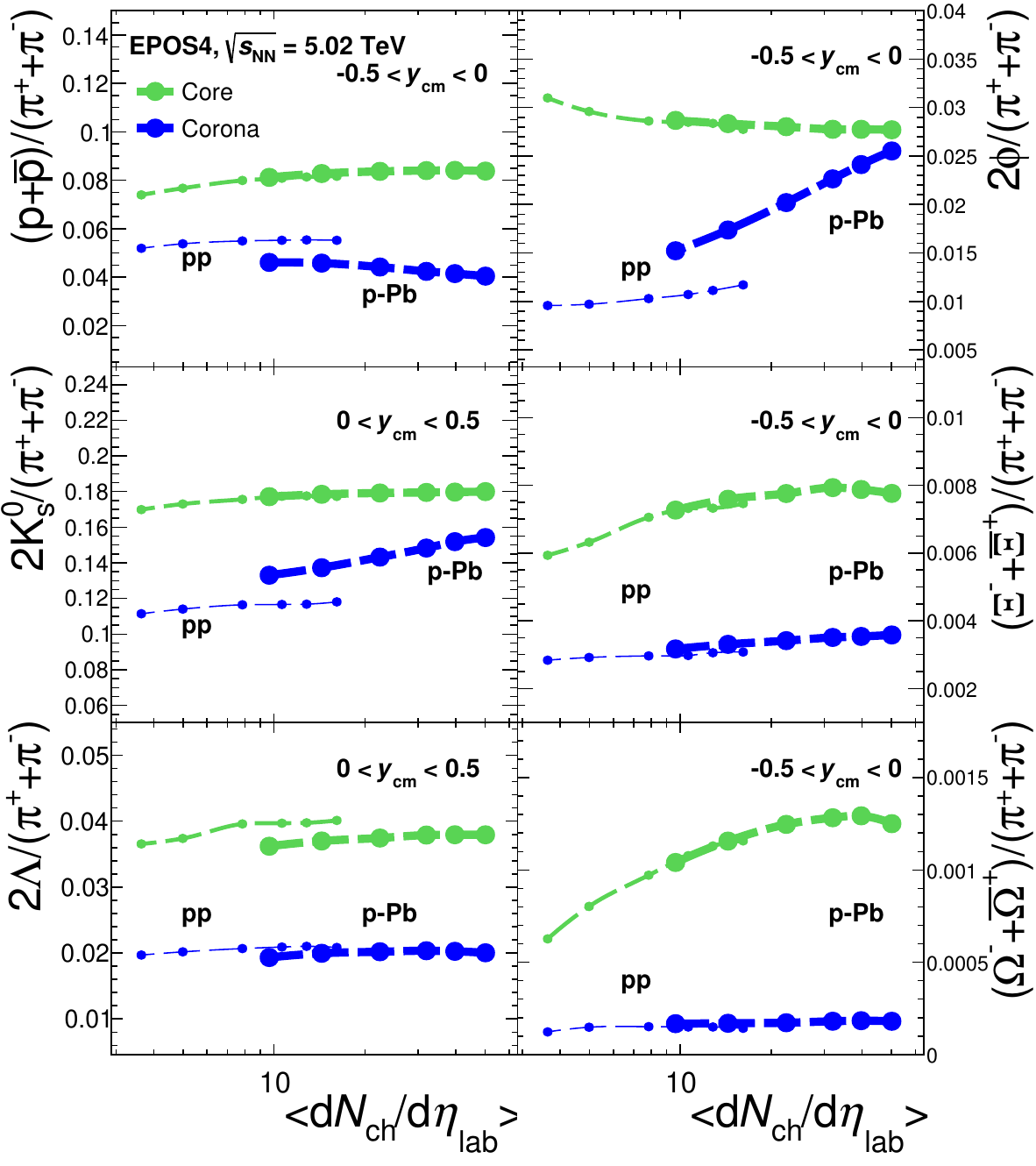}		
	\caption{(Color online)  Core-corona dependence of $p_{\rm{T}}$-integrated yield ratios of p, $K_{s}^{0}$, $\Lambda$, $\Xi^{-}$, $\Omega^{-}$, and $\phi$ to pions $(\pi^{-} + \pi^{+})$ as a function of $\langle \frac{dN_{ch}}{d\eta} \rangle$ in pp and p-Pb collisions at $\sqrt{s_{\rm NN}}$ = 5.02 TeV} 
	\label{SEwCoreCoronawithNch}%
\end{figure}

To investigate the relative production of strange hadrons and facilitate comparison with existing results from p–Pb collisions, yield ratios of various hadrons to pions are evaluated as a function of charged-particle multiplicity. Figure~\ref{SEwithNch} presents the $p_{\rm{T}}$-integrated yield ratios of p, $K_{s}^{0}$, $\Lambda$, $\Xi$, $\Omega$, and $\phi$ over pions, as obtained using the EPOS4 model in pp and p–Pb collisions at $\sqrt{s_{\mathrm{NN}}} = 5.02$ TeV.

A clear enhancement in the production of strange to non-strange hadrons is observed, which increases steadily with multiplicity from pp to p–Pb collisions. EPOS4 predictions are in reasonable agreement with experimental data from p–Pb collisions~\cite{Acharya_2019}, successfully capturing the increasing trend of strangeness enhancement with event multiplicity. This behavior in EPOS4 is attributed to its use of a microcanonical sampling procedure, which enforces strict energy and flavor conservation. This method suppresses the production of heavier or multi-strange particles in small systems. In contrast, earlier versions of EPOS employed a grand-canonical approach that yields flat particle ratios across multiplicities, regardless of system size. As shown in reference~\cite{EPOS4_pp_PbPb}, the discrepancy between microcanonical and grand-canonical predictions is most significant at low multiplicity, gradually diminishing as event multiplicity increases.

It is evident from Fig.~\ref{SEwithNch}, that the relative enhancement with multiplicity is most pronounced for multistrange baryons such as $\Omega$, compared to singly-strange hadrons like $\Lambda$ and $K_{s}^{0}$, with a steeper rise observed in pp collisions than in p–Pb. This supports the interpretation that the degree of enhancement is strongly correlated with a hadron's strangeness content, and appears to be governed primarily by final-state properties rather than the nature of the colliding system or energy. At high multiplicities, yield ratios tend to flatten, showing little or no further dependence on multiplicity.

It is also noteworthy that the final-state average charged-particle density, $\langle \frac{dN_{ch}}{d\eta} \rangle$, varies by up to two orders of magnitude from low-multiplicity pp to central p–Pb events. This large variation stems from fundamentally different underlying physics mechanisms: high-multiplicity events in pp collisions result mainly from multiple hard or semi-hard partonic scatterings, whereas in p–Pb collisions, similar multiplicities can arise from the superposition of multiple soft nucleon-nucleon interactions. Therefore, it is a non-trivial observation that particle yield ratios in pp and p–Pb collisions converge at the same $\langle \frac{dN_{ch}}{d\eta} \rangle$, suggesting that the final-state particle density may serve as a universal scaling variable across different collision systems.

Figure~\ref{SEwCoreCoronawithNch} provides further insight by illustrating the core–corona decomposition of strange hadron-to-pion ratios in EPOS4. In the core hadronization component, a hierarchical pattern of strangeness enhancement is observed—most pronounced for $\Omega$, followed by $\Xi$ and then $\Lambda$. 
In contrast, the corona contribution to strange hadron-to-pion ratios remains largely flat as a function of $\langle \frac{dN_{ch}}{d\eta} \rangle$ for most particle species, with notable exceptions being K$_{s}^{0}$ and the $\phi$ meson. As discussed earlier, this rise may stem from an increase in the effective string mass with event multiplicity, driven by a higher probability of hard interactions that facilitate s$\bar{\rm{s}}$ quark-pair production.

\subsection{Strangeness hierarchy}

\begin{figure}
	\centering 
 \includegraphics[width=0.48\textwidth, angle=0]{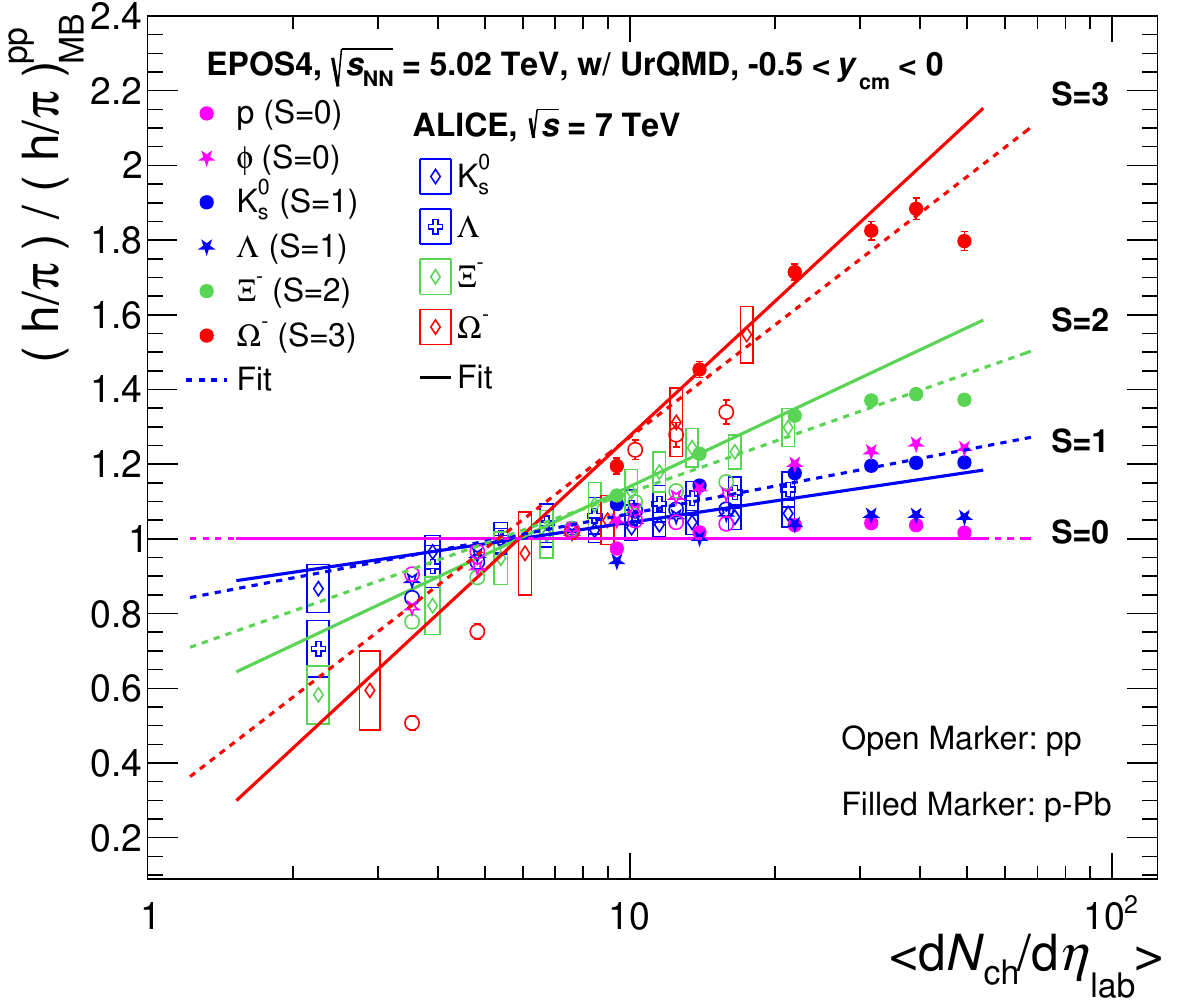}
	\caption{(Color online) Particle yield ratios for pp and p-Pb collisions at $\sqrt{s_{\rm NN}}$ = 5.02 TeV normalized to the values measured in the inclusive pp sample. Open markers with box error correspond to ALICE results and solid/open markers with vertical error bars (within the marker) represent EPOS4 results w/ UrQMD. Solid and dashed lines demonstrate fits to ALICE and EPOS4 results respectively, using Eq.1 (further details can be found in the text).} 
	\label{hStrangewNch}%
\end{figure}

Figure~\ref{hStrangewNch} presents the yield ratios of strange hadrons to pions, in pp and p-Pb collisions normalized to the corresponding values measured in the inclusive pp sample. A clear multiplicity-dependent enhancement is observed, which follows a hierarchy based on the strangeness content of the hadrons—most prominent for multi-strange baryons such as $\Omega$, followed by $\Xi$, $\Lambda$, and K$_{s}^{0}$. This trend highlights the progressive enhancement of strangeness production with increasing charged-particle multiplicity.

We have attempted to describe the observed behavior using the EPOS4 model by fitting the results presented in Fig.~\ref{hStrangewNch} with an empirical functional form:

\begin{equation}
\frac{\left\langle h / \pi \right\rangle}{\left\langle h / \pi \right\rangle^{\text{pp}}_{\text{MB}}} = 1 + a \, S^{b} \log \left[ \frac{\left\langle dN_{ch}/d\eta \right\rangle}{\left\langle dN_{ch}/d\eta \right\rangle^{\text{pp}}_{\text{MB}}} \right].
\label{placeholder_label}
\end{equation}

Here, S represents the number of strange or anti-strange valence quarks in the hadron, ($h / \pi)^{pp}_{MB}$ and $<dN_{ch}/d\eta>^{pp}_{MB}$ are the measured hadron-to-pion ratio and the charged-particle multiplicity density in pp collisions, respectively, and a and b are free parameters.
The fit describes the ALICE results \cite{pp_Nature} well, yielding a = 0.083 ± 0.006, b = 1.67 ± 0.09, with a $\chi ^{2}$/ndf of 0.66. Whereas, for EPOS4 fit parameters: a= 0.146 ± 0.019, b = 0.436 ± 0.186, with a $\chi ^{2}$/ndf of 51.44.

The EPOS4 model can qualitatively capture the hierarchy and scaling behavior for multi-strange baryons such as $\Omega$ and $\Xi$. However, some tension remains in the description of the $\phi$-meson and $\Lambda$. Although the $\phi$-meson is net-strangeness neutral, its behavior aligns with that of particles carrying $S=1$. In contrast, the $\Lambda$, despite being a strange baryon with $S=1$, appears to follow the trend expected for $S=0$ particles.

This behavior in the EPOS4 model can be interpreted as the result of an interplay between intrinsic strangeness production and the modification of hadron yields caused by post-hadronization re-scattering effects.

\subsection{Freeze out dynamics for pp and p-Pb events}

The evolution of the transverse momentum ($p_{\rm{T}}$) distribution with increasing event multiplicity in pp and p–Pb collisions exhibits a notable resemblance to the behavior seen in larger collision systems. In heavy-ion collisions, this trend is commonly attributed to the hydrodynamic radial expansion of the produced medium. Such collective effects can be quantitatively investigated using the Boltzmann-Gibbs Blast-Wave (BG-BW) model \cite{PhysRevC.48.2462}, which provides insights into the kinetic freeze-out conditions of the system.
Assuming a radially expanding thermal source characterized by a kinetic freeze-out temperature $T_{\text{kin}}$ and an average transverse flow velocity $\langle \beta_{T} \rangle$, the transverse momentum distribution of the emitted particles is described by the BG-BW functional form:
\begin{equation}
    \frac{1}{p_{\rm{T}}} \frac{dN}{dp_{\rm{T}}} \propto \int_0^R r \, dr \, m_T I_0 \left( \frac{p_{\rm{T}} \sinh \rho}{T_{\text{kin}}} \right) K_1 \left( \frac{m_T \cosh \rho}{T_{\text{kin}}} \right),
\end{equation}
where the radial flow profile $\rho$ is given by:
\begin{equation}
    \rho = \tanh^{-1} \beta_T = \tanh^{-1} \left( \left( \frac{r}{R} \right)^n \beta_s \right).
\end{equation}
Here, $m_T = \sqrt{p_{\rm{T}}^2 + m^2}$ denotes the transverse mass, $I_{0}$ and $K_{1}$ are modified Bessel functions, $r$ is the radial coordinate in the transverse plane, $R$ is the maximum radial extent of the fireball, $\beta_T(r)$ is the local transverse expansion velocity, $\beta_s$ is the velocity at the surface, and $n$ defines the velocity profile shape.

Typically, the $p_{\rm{T}}$ spectra of $\pi^{\pm}$, $K^{\pm}$, and $p/\bar{p}$ are simultaneously fitted using the BG-BW model to extract freeze-out parameters. In this study, we also include $K_{s}^{0}$ and $\Lambda/\bar{\Lambda}$ in the fit. However, multistrange hadrons are excluded, as imposing a common freeze-out condition on all species may not be valid; experimental observations suggest that multistrange hadrons may decouple from the medium earlier due to their smaller hadronic interaction cross-sections \cite{Xu_2002}.

\begin{figure}
	\centering 
    \includegraphics[width=0.48\textwidth, angle=0]{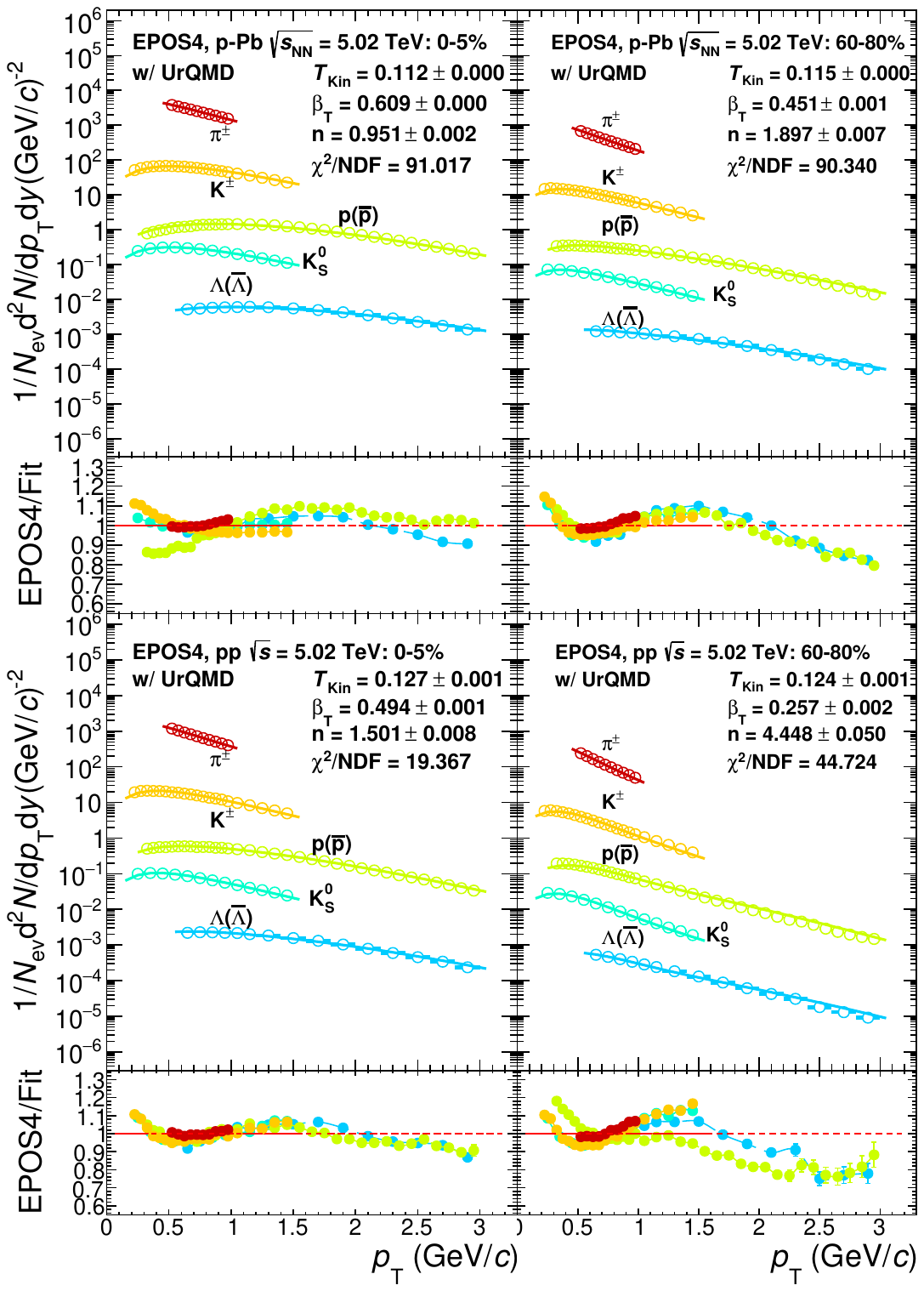}	
	\caption{(Color online) (Top) Simultaneous blast-wave fits to $\pi^{\pm}$, K$^{\pm}$, $K_s^0$, $p$($\bar{p}$), and $\Lambda/\bar{\Lambda}$ spectra in the fit ranges 0.5–1 GeV/$c$, 0.2–1.5 GeV/$c$, 0–1.5 GeV/$c$, 0.3–3 GeV/$c$, and 0.6–3 GeV/$c$, respectively, for pp and p–Pb collisions at $\sqrt{s_{\rm NN}} = 5.02$ TeV. Results are shown for the 0–5\% and 60–80\% multiplicity classes using event samples generated with EPOS4 w/ UrQMD. Solid lines represent the blast-wave fits to the corresponding spectra. (Bottom) Ratios of EPOS4 spectra to the blast-wave fits.} 
	\label{cBW_Combined}%
\end{figure}

The $p_{\rm{T}}$ ranges used in the fits are as follows: 0.5–1 GeV/$c$ for $\pi^{\pm}$, 0.2–1.5 GeV/$c$ for $K^{\pm}$, 0–1.5 GeV/$c$ for $K_s^0$, 0.3–3 GeV/$c$ for $p/\bar{p}$, and 0.6–3 GeV/$c$ for $\Lambda/\bar{\Lambda}$. These intervals are chosen to match the fit ranges employed in the ALICE experiment \cite{pikaprlamks0_alice}. The corresponding fits are shown in Fig.~\ref{cBW_Combined} for the 0–5\% and 60–80\% multiplicity classes of pp and p–Pb collisions from EPOS4 w/ UrQMD at $\sqrt{s_{\rm NN}} = 5.02$ TeV. Since the $\chi^{2}/\rm{NDF}$ values are relatively large, we additionally show the EPOS4-to-fit ratios in the bottom panels to better demonstrate the goodness of fit. Within the chosen $p_{\rm{T}}$ ranges, these ratios do not exceed 15–20\%.

\begin{figure}
	\centering 
    \includegraphics[width=0.48\textwidth, angle=0]{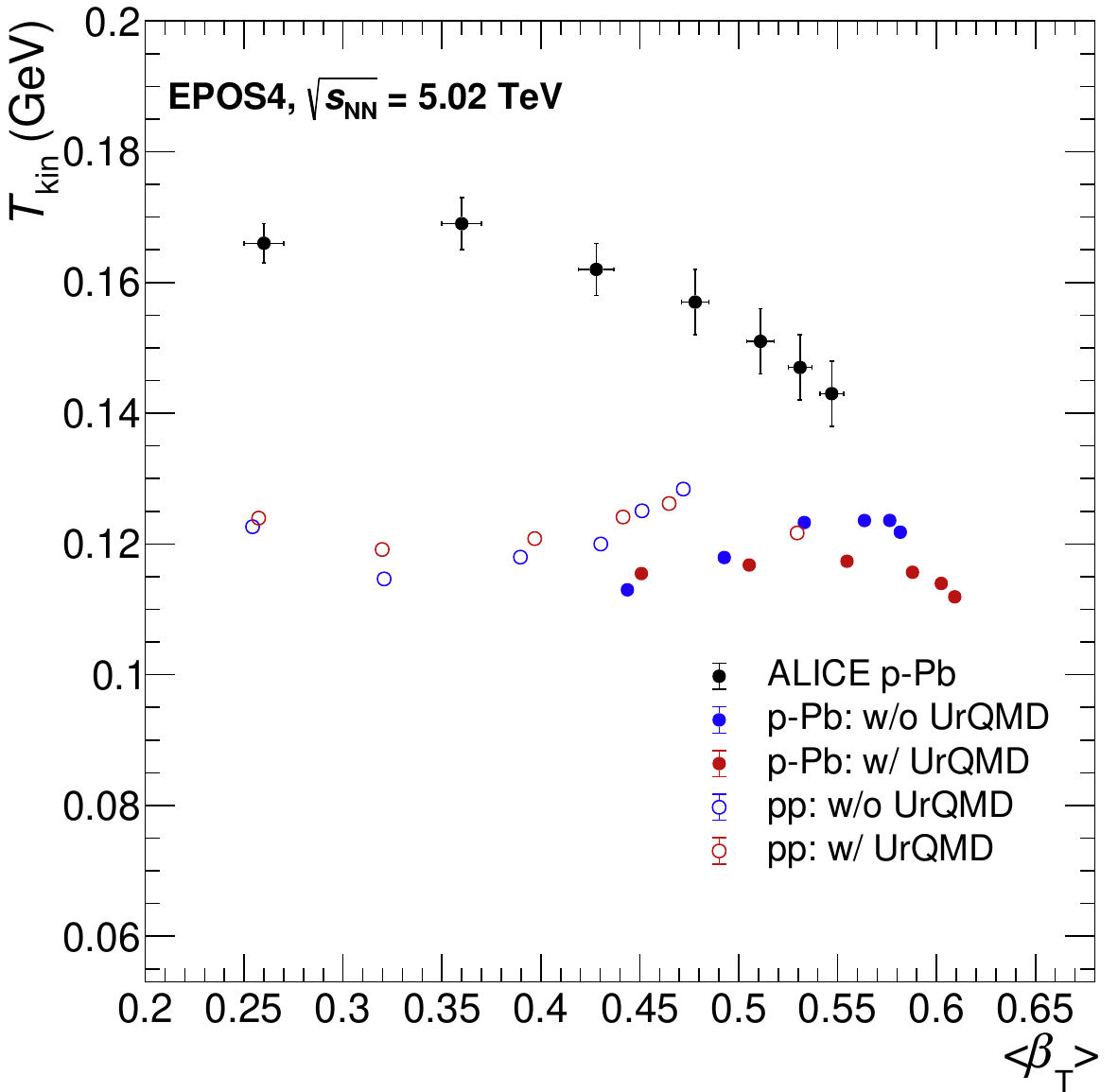}	
	\caption{(Color online) Results of blast-wave fits for pp, and p-Pb collisions at $\sqrt{s_{\rm NN}}$ = 5.02 TeV, compared to ALICE results \cite{pikaprlamks0_alice}. Charged particle multiplicity increases from left to right.} 
	\label{betaT}%
\end{figure}

\begin{figure*}[t]
	\centering 
    \includegraphics[width=0.95\textwidth, angle=0]{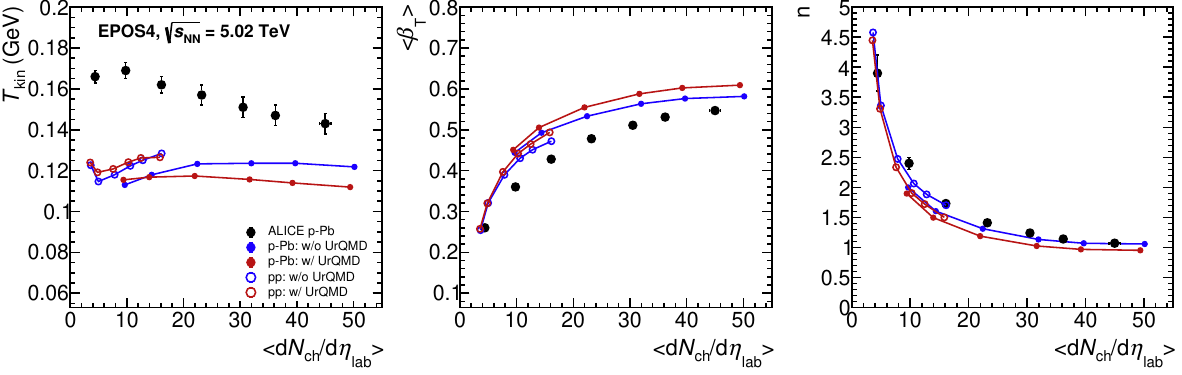}
	\caption{(Color online) Multiplicity dependence of blast-wave fit parameters: (left) kinetic freeze-out temperature $T_{\rm kin}$, (middle) average radial flow velocity $\langle \beta_{T} \rangle$, and (right) the velocity profile exponent $n$, obtained from simultaneous fits to $\pi^{\pm}$, K$^{\pm}$, $K_s^0$, $p$($\bar{p}$), and $\Lambda/\bar{\Lambda}$ spectra using EPOS4 w/ and w/o UrQMD. Results are compared with ALICE measurements \cite{pikaprlamks0_alice} in p–Pb collisions at $\sqrt{s_{\rm NN}} = 5.02$ TeV. } 
	\label{ParamwMult}%
\end{figure*}

Figure~\ref{betaT} displays the extracted $T_{\text{kin}}$ versus $\langle \beta_{T} \rangle$ from simultaneous BG-BW fits to experimental data for p–Pb collisions, as well as from EPOS4 model simulations for pp and p–Pb collisions at $\sqrt{s_{\rm NN}} = 5.02$ TeV. A clear anti-correlation between these parameters is observed: higher $T_{\text{kin}}$ corresponds to lower $\langle \beta_{T} \rangle$, and vice versa. A more detailed comparison between data and model predictions is presented in Fig.~\ref{ParamwMult}. The EPOS4-extracted $T_{\text{kin}}$ values are significantly lower than those from experimental data in similar multiplicity classes, while the $\langle \beta_{T} \rangle$ values tend to be higher.

In the case of p–Pb collisions, high-multiplicity event selection tends to bias the sample toward harder collisions \cite{Adam_2015_ALICE}, which may contribute to the larger $\langle \beta_{T} \rangle$ values obtained from the blast-wave fits. Moreover, EPOS4 simulations that include hadronic re-scattering exhibit even higher $\langle \beta_{T} \rangle$ values compared to simulations without re-scattering, indicating that hadronic re-scattering imparts additional radial flow to the system.

\section{Summary and conclusions}
The observation of collectivity and strangeness enhancement in small collision systems, such as proton-proton (pp) and proton-lead (p-Pb) collisions, challenges the conventional understanding of small system dynamics. Traditionally, these collisions were considered too small and short-lived to achieve thermalization, and thus, collective behavior and enhanced strangeness production were not anticipated. However, experimental data from high-energy collisions at the LHC and RHIC provide compelling evidence to the contrary. The data reveal clear signs of long-range correlations, even at very low multiplicities (10–20 per unit $\eta$) \cite{Acharya_2024}, as well as significant strangeness enhancement. These observations suggest the involvement of non-trivial particle production mechanisms that extend beyond conventional perturbative and non-perturbative QCD processes.

While initial-state models, particularly those within the framework of the Color Glass Condensate (CGC) effective field theory, can account for long-range correlations in small systems, the contribution of final-state interactions—particularly the strong interactions among the produced particles—cannot be neglected. These final-state interactions, modeled by hydrodynamic approaches, play a crucial role in explaining the observed correlations, flow patterns, and particle yields, especially in small systems like pp and p-Pb collisions. Moreover, in small systems where thermalization is not expected, the enhancement of strangeness production is particularly noteworthy. The levels of strange hadron production in high-multiplicity pp and p-Pb collisions have been found to be surprisingly comparable to those observed in larger systems. This phenomenon cannot be fully explained without assuming some form of thermalization or, at the least, the presence of significant final-state effects. Consequently, it is realized that models that combine initial-state saturation effects with final-state interactions are essential for capturing a broad spectrum of observed phenomena across systems of different sizes.

In this context, the EPOS model leverages a unified formalism within its core-corona framework to describe a wide range of phenomena from pp to p-A to A-A collisions. Taking advantage of this uniqueness in the model, we perform a detailed investigation of particle production in pp and p-Pb collisions at $\sqrt{s_{\rm NN}}$ = 5.02 TeV. Here we have studied $p_{\rm{T}}$-differential and $p_{\rm{T}}$-integrated yields of light, and (multi-)strange particles along with the particle ratios as a function of $p_{\rm{T}}$ and charged-particle multiplicity at mid-rapidity. Additionally, we investigate the individual contributions of the core and corona components to various physical observables.

In both pp and p-Pb collisions, the $p_{\rm{T}}$ spectra of all particle species become progressively harder with increasing average charged-particle multiplicity, $\langle \frac{dN{ch}}{d\eta} \rangle$. This hardening effect is more pronounced for heavier particles. The underlying reason is that the core contribution to particle production increases with multiplicity, leading to a stronger radial boost. Consequently, the spectral shapes of heavier particles are hardened more significantly compared to their lighter counterparts.
The EPOS4 model qualitatively reproduces the $p_{\rm{T}}$ spectra trends observed by ALICE but exhibits quantitative discrepancies that can go upto 40\%, depending on the particle species and multiplicity interval considered.

Regarding integrated yields, $\frac{dN}{dy}$, EPOS4 predicts a smooth and continuous increase with $\langle \frac{dN_{ch}}{d\eta} \rangle$ from pp to p-Pb collisions, reasonably capturing the non-linear trend seen in the data with some deviations for specific particles, particularly the $\phi$-meson and $\Lambda$. EPOS4 tends to overestimate the $\phi$-meson yields while underestimating those of the $\Lambda$. For the $\phi$-meson, the increasing contribution from the corona component with multiplicity, unlike other strange hadrons, could be a reason for the overestimation. In the case of the $\Lambda$, baryon–antibaryon annihilation effects may suppress its yield. Since annihilation cross-sections are largely model-dependent, better tuning of the annihilation parameters may improve agreement with the data.

The comparison of average transverse momentum, $\langle p_{\rm{T}} \rangle$, as a function of multiplicity reveals reasonable agreement between EPOS4 predictions and experimental data for most particle species, except for pions and multi-strange hadrons such as the $\Omega^{-}$.
Generally, $\langle p_{\rm{T}} \rangle$ increases monotonically with particle mass, consistent with collective flow effects. However, experimental observations exhibit a violation of this mass ordering, exemplified by the $\langle p_{\rm{T}} \rangle_{\phi}$ being higher than proton and $\Lambda$ baryons. In contrast, EPOS4 does not show such an overall violation. When the model’s output is decomposed into core and corona components, the core—which represents the thermalized, collectively flowing medium—displays a clear mass ordering of $\langle p_{\rm{T}} \rangle$, while the corona component exhibits deviations from this pattern. Due to the dominant contribution of the core, the combined $\langle p_{\rm{T}} \rangle$  effectively masks the corona-induced violations. It is important to emphasize that a model must simultaneously describe both particle yields and $\langle p_{\rm{T}} \rangle$. This is because how the available collision energy is allocated between producing new particles (through hadronization processes) and providing kinetic energy to these particles crucially constrains the underlying particle production mechanisms. In the context of the core-corona approach in EPOS4, the balance between the energy utilized for particle production and for imparting transverse motion in both components can differ, impacting the final state observables like particle $p_{\rm{T}}$-distributions, yields and particle ratios.
 
Particle yield ratios studied as a function of $p_{\rm{T}}$ and event multiplicity offer further insight on the underlying hadronization mechanisms involved. At low multiplicities, hadron production is often dominated by vacuum-like fragmentation processes, while at high multiplicities, modifications to hadronization may occur due to the formation of a dense medium or increased parton interactions. Therefore, observing systematic changes in particle yield ratios as a function of both $p_{\rm{T}}$ and multiplicity helps to reveal the transition between these regimes. In EPOS4, the particle yield ratios, p/$\pi$ and $\Lambda/K_{s}^{0}$ exhibit characteristic enhancement at intermediate $p_{\rm{T}}$, consistent with the collective flow picture in the model. The experimental data show very good agreement with EPOS4 predictions for the $\Lambda/K_{s}^{0}$ ratios but overestimates the same for p/$\pi$. This discrepancy may be attributed to an overestimation of the radial flow velocity in EPOS4, suggesting that a better tuning of the input parameters that drives the flow, such as the initial energy density profile, initial flow velocity, choice of equation of states (EOS) or the values of $\eta/s$, could improve the description of baryon-to-meson ratios further, especially for protons. Interestingly, despite the mass difference between the K and $\phi$ mesons being comparable to that between $\Lambda$ and  $K_{s}^{0}$, the $\phi/K$ ratio does not display any enhancement with $p_{\rm{T}}$. This difference highlights that the particle production cannot be fully explained by simple mass scaling, indicating that other factors like quark composition, baryon number, etc  has a role to play. Within the EPOS4 framework, it has been observed that non-thermal production of the $\phi$-meson from the corona region increases significantly with event multiplicity. The corona represents regions in the collision with lower density and less collective behavior, where particle production follows vacuum-like fragmentation rather than thermalized medium emission. The increase in non-thermal components could dilute any collective flow effects in  $\phi$-meson yields. Furthermore, re-scattering effects on high-$p_{\rm{T}}$ $\phi$-meson are supposedly small because of its low hadronic interaction cross sections allowing it to escape the medium largely unmodified. 

EPOS4, through its implementation of a microcanonical approach to core hadronization, successfully captures the observed trend of strangeness enhancement in strange-to-pion yield ratios as a function of event multiplicity. The microcanonical approach enforces strict conservation of energy, momentum, and quantum numbers (such as baryon number, strangeness, and charge) within localized regions of the hypersurface formed during the hadronization. This leads to more constrained modeling of particle production in small systems compared to the grand-canonical approach used in earlier versions of EPOS (e.g., EPOS3).

Specifically, the microcanonical framework naturally accounts for the suppression of strangeness in low-multiplicity events due to limited available phase space and conservation constraints. As multiplicity increases—corresponding to an effectively larger available phase space—the constraints are relaxed, allowing for enhanced production of strange hadrons relative to pions, gradually converging toward the grand-canonical expectation at large multiplicities. Remarkably, particle production from microcanonical sampling of the core alone qualitatively reproduces the experimentally observed strangeness enhancement trend without introducing additional parameters.

EPOS4 also includes the flexibility to further control yields of multi-strange hadrons by restricting microcanonical sampling to smaller hypersurface elements, achieved by dividing the entire hypersurface into smaller 
$\Delta\eta$ regions. This procedure causes additional suppression of multi-strange particle yields in particular. Although counterintuitive, the best agreement with experimental data is achieved by applying microcanonical sampling over the entire hypersurface rather than limiting it to smaller regions.

EPOS4 also reproduces the hierarchical nature of strangeness production depending on the strangeness content, with notable exceptions being the $\phi$-meson and $\Lambda$ baryon. Although $\phi$-meson carries zero net strangeness, its behavior appears consistent with an effective strangeness S=1, whereas, $\Lambda$ baryon aligns more closely with S=0. For $\phi$-meson this discrepancy may be attributed to the increasing contribution of non-thermal component with multiplicity, while for the $\Lambda$,  it could result from an overestimation of the baryon–antibaryon annihilation cross section in the model.

Finally, the kinetic freeze-out parameters, T$_{kin}$ and $\beta$, extracted from combined Blast-Wave fits to $\pi$, $K$, $p/\bar{p}$, $K_{s}^{0}$ and $\Lambda$ $p_{\rm{T}}$-spectra generated by EPOS4, deviate significantly from experimental data. The extracted values of the radial flow parameter, $\beta$, are larger than that of the data, consistent with the model’s overestimation of the pion $\langle p_{\rm{T}} \rangle$.

In summary, EPOS4 provides a comprehensive framework that successfully reproduces many qualitative features of hadron production and strangeness enhancement across various multiplicities and transverse momenta. Its microcanonical approach to core hadronization offers a more realistic treatment of conservation laws in small systems, leading to improved modeling of strange hadron yields. However, challenges remain in achieving quantitative agreement with experimental data. Future developments, including improved balancing of core and corona contributions, different freeze-out conditions for multi-strange hadrons, and implementation of finite strangeness correlation volumes,  may enhance the model's quantitative predictability and agreement with experimental measurements. Such advancements will indeed be crucial for deepening our understanding of the complex dynamics that govern particle production in high-energy collisions.

\section*{Acknowledgements}
The authors thank Dr. Klaus Warner for providing us with the EPOS4 model. 
The authors are thankful to the members of the grid computing team of VECC and the cluster computing team of the Department of Physics, Jadavpur University for providing uninterrupted facilities for event generation and analyses. 
One of the authors, HK, acknowledes Dr. Somnath Kar for valuable discussion on usage of EPOS4.
We also gratefully acknowledge the financial help from the DST-GOI under the scheme “Mega facilities for basic science research” [Sanction Order No. SR/MF/PS-02/2021-Jadavpur (E-37128) dated December 31, 2021].

\nocite{*}

\bibliography{apssamp} 

\end{document}